\documentclass[english,onecolumn]{IEEEtran}
\usepackage[T1]{fontenc}
\usepackage[latin9]{inputenc}
\usepackage{amsmath}
\usepackage{amsthm}
\usepackage{amssymb}
\usepackage{graphicx}
\usepackage{esint}

\makeatletter
\theoremstyle{plain}
\newtheorem{thm}{\protect\theoremname}
\theoremstyle{remark}
\newtheorem{rem}[thm]{\protect\remarkname}
\theoremstyle{plain}
\newtheorem{cor}[thm]{\protect\corollaryname}
\theoremstyle{plain}
\newtheorem{prop}[thm]{\protect\propositionname}

\usepackage{mathrsfs}
\usepackage{algorithm,algpseudocode}

\usepackage{colortbl}
\definecolor{lightgray}{rgb}{0.9,0.9,0.9}
\definecolor{lightred}{rgb}{1,0.8,0.8}
\definecolor{lightgreen}{rgb}{0.6,1,0.6}
\definecolor{lightyellow}{rgb}{1,1,0.5}
\definecolor{lightgrey}{rgb}{0.8,0.8,0.8}

\allowdisplaybreaks[1]

\makeatother

\usepackage{babel}
\providecommand{\corollaryname}{Corollary}
\providecommand{\propositionname}{Proposition}
\providecommand{\remarkname}{Remark}
\providecommand{\theoremname}{Theorem}

\begin{document}
\title{Undecidability of Network Coding, Conditional Information Inequalities,
and Conditional Independence Implication}
\author{Cheuk Ting Li\\
Department of Information Engineering, The Chinese University of Hong
Kong\\
Email: ctli@ie.cuhk.edu.hk}
\maketitle
\begin{abstract}
We resolve three long-standing open problems, namely the (algorithmic)
decidability of network coding, the decidability of conditional information
inequalities, and the decidability of conditional independence implication
among random variables, by showing that these problems are undecidable.
The proof utilizes a construction inspired by Herrmann's arguments
on embedded multivalued database dependencies, a network studied by
Dougherty, Freiling and Zeger, together with a novel construction
to represent group automorphisms on top of the network.
\end{abstract}

\begin{IEEEkeywords}
Network coding, conditional information inequalities, conditional
independence implication, index coding, Turing undecidability.
\end{IEEEkeywords}

\medskip{}

\section{Introduction}

Network coding \cite{ahlswede2000network,li2003linear} is a setting
in which each node in a network can perform encoding and decoding
operations, in order to transmit some messages through the network.
There are numerous research works on algorithms for network coding,
e.g. \cite{jaggi2005polynomial,harvey2005deterministic,li2005achieving,ho2006random}.
However, it was shown that various problems about network coding are
NP-hard, e.g. \cite{lehman2005network,langberg2006encoding,yao2009network,langberg2011hardness}. 

It was uncertain if network coding is even decidable, that is, if
there exists an algorithm that can determine whether a network is
solvable (i.e., admits a coding scheme satisfying the decoding constraints)
\cite{lehman2005network,dougherty2011network}. Network coding would
be decidable if there is a computable upper bound on the alphabet
size, as observed by Rasala Lehman \cite{lehman2005network}. For
partial undecidability results, K{\"u}hne and Yashfe \cite{kuhne2019representability}
proved that determining whether a network admits a vector linear network
code is undecidable. A potential approach to show undecidability was
proposed by Dougherty \cite{dougherty2009undecidable}, which involves
a reduction from Rhodes' problem, i.e., the identity problem for finite
groups, which is conjectured to be undecidable \cite{albert1992undecidability}.
If a subset of the messages and edges have a fixed size different
from the common size of other messages/edges, then Li \cite{li2021netcode}
showed that the solvability of a network is undecidable. For other
related works and discussions on this open problem, refer to \cite{cannons2006network,dougherty2008linear,langberg2009multiple,dougherty2011network,bassoli2013network,huang2013secure,gomez2014network}.

A related problem is the conditional independence implication problem
\cite{dawid1979conditional,spohn1980stochastic,mouchart1984note,pearl1987graphoids},
which is to decide whether a statement on the conditional independence
among some random variables follows from a list of other such statements.
Pearl and Paz \cite{pearl1987graphoids} introduced a set of axioms,
called the semi-graphoid axioms, as a proposed axiomization of conditional
independence. This set of axioms was shown to be incomplete by Studen\'y
\cite{studeny1989multiinformation,studeny1990conditional}. Partial
decidability/undecidability results has been obtained in \cite{geiger1991axioms,geiger1993logical,geiger1999quantifier,niepert2012logical,gyssens2014completeness,hannula2019facets,khamis2020decision,li2021undecidability}.
In particular, it was noted by Geiger and Meek \cite{geiger1999quantifier}
and Niepert \cite{niepert2012logical} that if all random variables
have bounded cardinalities, then the implication problem is decidable.
If only a subset of the random variables have bounded cardinalities,
then Li \cite{li2021undecidability} has shown that the problem is
undecidable. If we consider the whole first-order theory of random
variables with the conditional independence relation (instead of only
the implication problem), then this theory is undecidable \cite{li2021first}.
Nevertheless, the decidability of the conditional independence implication
problem remained open \cite{geiger1999quantifier,wong2000implication,niepert2010logical,niepert2013conditional,koehler2014saturated,khamis2020decision}.

Another closely related problem is conditional information inequalities
\cite{zhang1997non,yeung1997framework,zhang1998characterization},
which is to decide whether a linear inequality involving entropy terms
among some random variables follows from a list of other such inequalities.
This problem generalizes the conditional independence implication
problem since the conditional independence $X\perp Y|Z$ can be expressed
as $I(X;Y|Z)\le0$. The connection to network coding was studied in
\cite{yeung2008information,chan2008dualities,yan2012implicit}. Zhang
and Yeung showed the first conditional non-Shannon-type inequality
(i.e., cannot be proved using only the fact $I(X;Y|Z)\ge0$) in \cite{zhang1997non},
and the first unconditional non-Shannon-type inequality in \cite{zhang1998characterization}.
Also see \cite{makarychev2002new,dougherty2006six,matus2007infinitely,xu2008projection,dougherty2011non,kaced2013conditional}
for more non-Shannon-type inequalities. While Shannon-type inequalities
can be verified algorithmically \cite{yeung1997framework,yeung1996itip},
there are also algorithms capable of verifying some non-Shannon-type
information inequalities \cite{xu2008projection,gurpinar2019use,ho2020proving,li2021automatedisit}.
If we allow affine (instead of only linear) inequalities, then it
was shown by Li \cite{li2021undecidability} that the problem is undecidable.
The decidability of linear conditional (and unconditional) information
inequalities remained open \cite{gomez2014network,gomez2017defining,khamis2020decision,khamis2021bag,yeung2021machine}. 

In this paper, we resolve these open problems by proving the undecidability
of the three aforementioned problems. The proof of the undecidability
of conditional independence implication (Theorem \ref{thm:undecide_ci})
is inspired by Herrmann's proof \cite{herrmann1995undecidability,herrmann2006corrigendum}
on the undecidability of embedded multivalued database dependencies
(EMVD) \cite{fagin1977multivalued}, which uses a reduction from the
uniform word problem for finite semigroups/monoids. While EMVD shares
several similarities with probabilistic conditional independence among
random variables (e.g. they both satisfy the semi-graphoid axioms
\cite{pearl1987graphoids}), the valid implications in probabilistic
conditional independence is neither a subset nor a superset of the
valid implications in EMVD \cite{studeny1990conditional}. Therefore,
the undecidability of conditional independence implication is not
a direct corollary of the undecidability of EMVD, and arguments specific
to random variables are needed to show the undecidability of conditional
independence implication.

The proof of the undecidability of network coding (Theorem \ref{thm:netcode})
utilizes a reduction from the uniform word problem for finite groups,
where the groups are embedded as subgroups of the automorphism group
of an abelian group. We utilizes a network studied in \cite{dougherty2006unachievability},
which captures the structure of an abelian group, together with a
novel construction to represent automorphisms of the abelian group
using subnetworks.

As a result, the minimum alphabet size needed to solve a network can
be uncomputably large, i.e., not upper-bounded by any computable function
(Corollary \ref{cor:computable}). This is a direct corollary of the
undecidability of network coding and the observation in \cite{lehman2005network}
that network coding would be decidable if there is a computable upper
bound on the alphabet size. Comparing to the result in \cite{lehman2005networkmodel}
which showed the existence of networks which are solvable only for
an alphabet size double exponential in the number of nodes and messages,
in this paper we show that there are solvable networks whether even
double exponential (or triple exponential, etc) would not be sufficient. 

Another corollary is that network coding for multiple unicast networks
\cite{li2004network} (i.e., each source message is available to one
source node and demanded by one receiver node) is also undecidable,
due to the result in \cite{dougherty2006nonreversibility} that a
general network coding problem can be reduced to a multiple unicast
setting. Index coding \cite{bar2011index,lubetzky2009nonlinear} is
undecidable as well, which is due to the equivalence between network
coding and index coding \cite{el2010index,effros2015equivalence}.

The paper is organized as follows. In Section \ref{sec:fano}, we
introduce the Fano-non-Fano condition. In Section \ref{sec:monoid},
we completes the proof of the undecidability of conditional independence
implication by showing a reduction from the uniform word problem for
finite monoids. In Section \ref{sec:network}, we prove the undecidability
of network coding. 

\medskip{}

\begin{rem}
The proof of the undecidability of conditional independence implication
(Theorem \ref{thm:undecide_ci}) originated as an attempt to adapt
the proof in \cite{herrmann1995undecidability} (which concerns EMVD
instead of random variables) into an argument on random variables.
While the final proof of Theorem \ref{thm:undecide_ci} in Sections
\ref{sec:fano} and \ref{sec:monoid} still follows the high-level
approach in \cite{herrmann1995undecidability} (i.e., showing a reduction
from the uniform word problem for finite semigroups/monoids, via an
embedding into the endomorphism monoid of an abelian group), the details
became rather different from \cite{herrmann1995undecidability}, and
many of the proof techniques in this paper are novel. The parts that
are similar to \cite{herrmann1995undecidability} are marked explicitly. 

The constructions in Section \ref{sec:network} for the undecidability
of network coding, except the use of the network studied in \cite{dougherty2006unachievability},
are novel. Due to the design constraints of a network coding setting
(network coding is more restrictive than conditional independence
in the sense that some conditional independence relations cannot be
enforced by network coding), we require a reduction from the uniform
word problem for finite groups \cite{slobodskoi1981undecidability},
instead of the uniform word problem for finite semigroups/monoids
as in \cite{dougherty2006unachievability}.
\end{rem}
\medskip{}

\begin{rem}
The main difference between the proof of the undecidability of network
coding (Theorem \ref{thm:netcode}) in this paper and the approach
proposed by Dougherty \cite{dougherty2009undecidable} is that \cite{dougherty2009undecidable}
attempts to use edges to represent words and identities in groups
(an identity is in the form $w\equiv e$ where $w$ is a word, that
is, the equality $w=e$ is true for all substitution of letters in
$w$ by group elements), whereas we use edges to represent elements
in groups. Therefore, the approach in \cite{dougherty2009undecidable}
relies on the identity problem for finite groups (i.e., whether a
list of identities implies another identity), which is not known to
be decidable or undecidable \cite{albert1992undecidability}. This
contributes to one of the gaps in the approach in \cite{dougherty2009undecidable}.
On the other hand, the approach in this paper relies on the uniform
word problem for finite groups, which is known to be undecidable \cite{slobodskoi1981undecidability}.
\end{rem}
\medskip{}

\subsection*{Notations}

Throughout this paper, all random variables are assumed to have finite
support (i.e., finite random variables). The condition that two random
variables $X,Y$ are independent is written as $X\perp Y$. The condition
that two random variables $X,Y$ are independent conditional on $Z$
is written as $X\perp Y|Z$. For random variables $X,Y$, we use juxtaposition
$XY$ to denote the joint random variable $(X,Y)$. For a sequence
of random variables $X_{1},\ldots,X_{k}$ and a set $\mathcal{U}\subseteq\{1,\ldots,k\}$,
write $X_{\mathcal{U}}=(X_{u_{1}},\ldots,X_{u_{|\mathcal{U}|}})$,
where $u_{1},\ldots,u_{|\mathcal{U}|}$ are the elements of $\mathcal{U}$
in ascending order. When we write $X=Y$ for random variables $X,Y$,
this means $X=Y$ holds with probability $1$. The logical conjunction
(i.e., ``AND'') between two statements $P,Q$ is denoted as $P\wedge Q$.
The logical conjunction between $P_{1},\ldots,P_{n}$ is denoted as
$\bigwedge_{i=1}^{n}P_{i}$.

\section{The Fano-non-Fano Condition\label{sec:fano}}

We will prove the first main result in this paper about conditional
independence implication.
\begin{thm}
\label{thm:undecide_ci}The following problem is undecidable: Given
$k,l\in\mathbb{Z}_{>0}$, $\mathcal{U}_{i},\mathcal{V}_{i},\mathcal{W}_{i}\subseteq\{1,\ldots,k\}$
for $i=0,\ldots,l$ satisfying $\mathcal{U}_{i}\cap\mathcal{V}_{i}=\mathcal{U}_{i}\cap\mathcal{W}_{i}=\mathcal{V}_{i}\cap\mathcal{W}_{i}=\emptyset$,
determine whether the implication
\[
\bigg(\bigwedge_{i=1}^{l}X_{\mathcal{U}_{i}}\perp X_{\mathcal{V}_{i}}|X_{\mathcal{W}_{i}}\bigg)\;\to\;X_{\mathcal{U}_{0}}\perp X_{\mathcal{V}_{0}}|X_{\mathcal{W}_{0}}
\]
holds for all jointly-distributed random variables $X_{1},\ldots,X_{k}$
with finite support.\footnote{Theorem \ref{thm:undecide_ci} continues to hold if $X_{1},\ldots,X_{k}$
are discrete random variables (with finite or countably infinite support).
This is because if some discrete random variables satisfy the $\mathrm{tri}$
condition \eqref{eq:triple}, then they must be finite.}
\end{thm}
\medskip{}

Since the problem of conditional information inequalities is a generalization
of the conditional independence implication problem, it is undecidable
as well. We state the result formally as follows. For a sequence of
finite random variables $X_{1},\ldots,X_{k}$, its entropic vector
\cite{zhang1997non} is defined as $\mathbf{h}(X_{1},\ldots,X_{k})=\mathbf{h}\in\mathbb{R}^{2^{k}-1}$,
where the entries of $\mathbf{h}$ are indexed by nonempty subsets
of $\{1,\ldots,k\}$, and $\mathbf{h}_{\mathcal{S}}:=H(X_{\mathcal{S}})$
(where $\mathcal{S}\subseteq\{1,\ldots,k\}$) is the joint entropy
of $\{X_{i}\}_{i\in\mathcal{S}}$. The following is a direct corollary
of Theorem \ref{thm:undecide_ci} and the fact that $X\perp Y|Z$
$\Leftrightarrow$ $-I(X;Y|Z)=H(X,Y,Z)+H(Z)-H(X,Z)-H(Y,Z)\ge0$.
\begin{cor}
\label{cor:entropy}The following problem is undecidable: Given $k\in\mathbb{Z}_{>0}$,
$\mathbf{a},\mathbf{b}\in\mathbb{Q}^{2^{k}-1}$, determine whether
the implication
\[
\mathbf{a}^{T}\mathbf{h}(X_{1},\ldots,X_{k})\stackrel{(i)}{\ge}0\;\to\;\mathbf{b}^{T}\mathbf{h}(X_{1},\ldots,X_{k})\stackrel{(ii)}{\ge}0
\]
holds for all jointly-distributed random variables $X_{1},\ldots,X_{k}$
with finite support.\footnote{Corollary \ref{cor:entropy} continues to hold if $X_{1},\ldots,X_{k}$
are discrete random variables (with finite or countably infinite support)
with finite entropy. } The problem remains undecidable if either one or both of the ``$\ge$''
marked with (i) and (ii) are replaced by ``$=$''.
\end{cor}
\medskip{}

The proof of Theorem \ref{thm:undecide_ci} is divided into Sections
\ref{sec:fano} and \ref{sec:monoid}. We begin with introducing some
notations. The condition that $X,Y,Z$ are mutually independent is
written as
\[
X\perp Y\perp Z\,\Leftrightarrow\,X\perp Y\,\wedge\,XY\perp Z.
\]
The condition that $Y$ contains no more information than $X$, i.e.,
$Y$ is a function of $X$, is written as
\[
Y\stackrel{\iota}{\le}X\;\Leftrightarrow\;Y\perp Y|X.
\]
Also, we write 
\[
X\stackrel{\iota}{=}Y\;\Leftrightarrow\;X\stackrel{\iota}{\le}Y\,\wedge\,Y\stackrel{\iota}{\le}X.
\]

We use the fact in \cite{zhang1997non} that if $Y_{1},Y_{2},Y_{3}$
satisfy that any one is a function of the other two, and they are
pairwise independent, then $Y_{1}$ is uniformly distributed over
its support (also true for $Y_{2},Y_{3}$) and $Y_{1},Y_{2},Y_{3}$
have the same cardinality. Also see the coordinatization via 3-net
in \cite{keedwell2015latin,herrmann1995undecidability}. Define the
predicate $\mathrm{tri}(Y_{1},Y_{2},Y_{3})$ over the variables $Y_{1},Y_{2},Y_{3}$
as
\begin{align}
 & \!\!\!\!\mathrm{tri}(Y_{1},Y_{2},Y_{3}):\nonumber \\
 & Y_{1}\stackrel{\iota}{\le}Y_{2}Y_{3}\,\wedge\,Y_{2}\stackrel{\iota}{\le}Y_{1}Y_{3}\,\wedge\,Y_{3}\stackrel{\iota}{\le}Y_{1}Y_{2}\nonumber \\
 & \wedge\,Y_{1}\perp Y_{2}\,\wedge\,Y_{1}\perp Y_{3}\,\wedge\,Y_{2}\perp Y_{3}.\label{eq:triple}
\end{align}
One example of $Y_{1},Y_{2},Y_{3}$ satisfying $\mathrm{tri}(Y_{1},Y_{2},Y_{3})$
is that $Y_{1},Y_{2}\sim\mathrm{Unif}(\mathcal{Y})$ i.i.d., and $Y_{3}=Y_{1}+Y_{2}$,
where $\mathcal{Y}$ is a finite abelian group.

\begin{figure}
\begin{centering}
\includegraphics[scale=0.5]{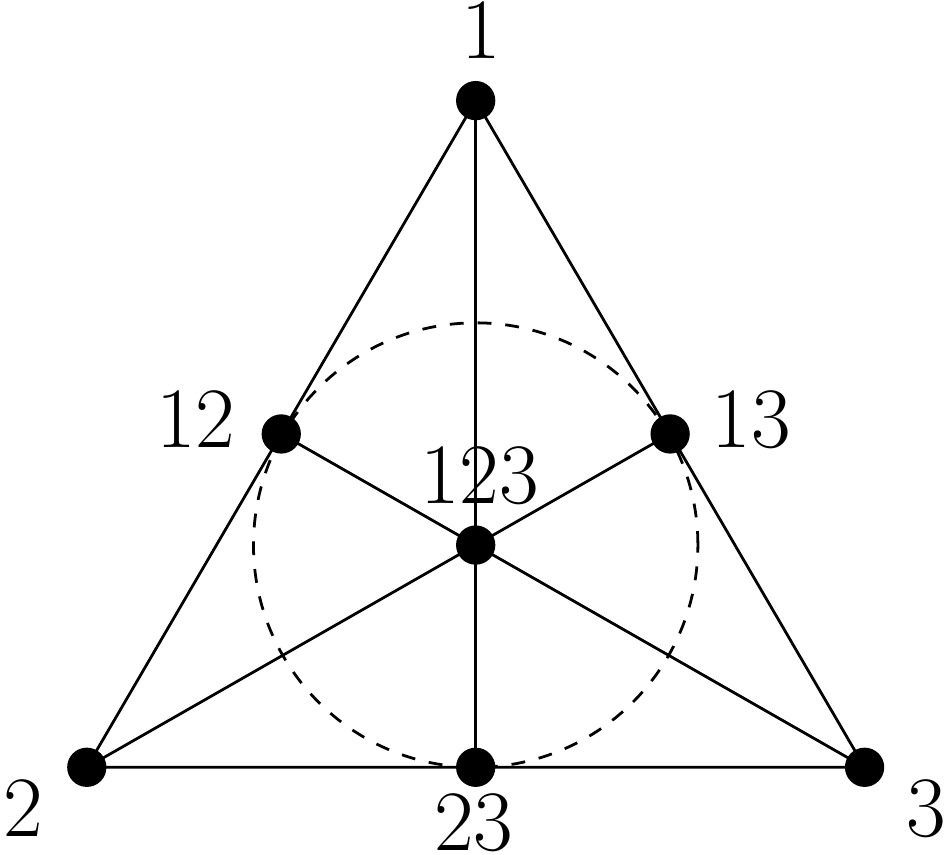}
\par\end{centering}
\caption{\label{fig:fano}The non-Fano matroid (solid lines only) and the Fano
matroid (solid and dotted lines). In the non-Fano matroid, a set of
size $3$ is dependent if and only if it is one of the solid lines.
In the Fano matroid, a set of size $3$ is dependent if and only if
it is one of the solid lines, or it is the dotted circle.}
\end{figure}

Our construction requires the Fano matroid and the non-Fano matroid
\cite{oxley2006matroid}. Let $\mathcal{E}=\{1,2,3,12,13,23,123\}$
be the ground set. In the non-Fano matroid \cite{oxley2006matroid,dougherty2007networks},
the set of dependent sets of size $3$ is given by
\begin{align*}
\mathcal{D}_{\mathrm{N}} & :=\big\{\{1,2,12\},\,\{1,3,13\},\,\{2,3,23\},\\
 & \;\;\;\;\;\;\;\{1,23,123\},\,\{2,13,123\},\,\{3,12,123\}\big\},
\end{align*}
which are the solid lines in Figure \ref{fig:fano}. In the Fano matroid
\cite{oxley2006matroid}, the set of dependent sets of size $3$ is
given by
\[
\mathcal{D}_{\mathrm{F}}:=\mathcal{D}_{\mathrm{N}}\cup\{\{12,23,13\}\},
\]
which are the solid lines together with the dotted circle in Figure
\ref{fig:fano}. Also write
\[
\mathcal{I}_{\mathrm{F}}:=\left\{ \{i,j,k\}\in2^{\mathcal{E}}:\,|\{i,j,k\}|=3\,\wedge\,\{i,j,k\}\notin\mathcal{D}_{\mathrm{F}}\right\} 
\]
for the set of independent sets of size $3$ in the Fano matroid. 

The \emph{Fano-non-Fano condition} on the seven random variables $A_{1},A_{2},A_{3},A_{12},A_{13},A_{23},A_{123}$
is defined as
\begin{align}
 & \mathrm{fnf}(A_{1},A_{2},A_{3},A_{12},A_{13},A_{23},A_{123}):\nonumber \\
 & \bigwedge_{\{i,j,k\}\in\mathcal{D}_{\mathrm{N}}}\mathrm{tri}(A_{i},A_{j},A_{k})\,\wedge\,\bigwedge_{\{i,j,k\}\in\mathcal{I}_{\mathrm{F}}}A_{i}\perp A_{j}\perp A_{k},\label{eq:fnf}
\end{align}
i.e., we enforce the dependent sets of size $3$ in the non-Fano matroid,
and the independent sets of size $3$ in the Fano matroid. Note that
$A_{12}\neq(A_{1},A_{2})$, and we treat the subscript $12$ as the
concatenation of two symbols $1$ and $2$ (or simply the number $12$)
instead of the set $\{1,2\}$. 

Note that $\bigwedge_{\{i,j,k\}\in\mathcal{I}_{\mathrm{F}}}A_{i}\perp A_{j}\perp A_{k}$
in \eqref{eq:fnf} can be replaced simply by $A_{1}\perp A_{2}\perp A_{3}$,
as shown below.
\begin{prop}
\label{prop:fnf_equiv}The Fano-non-Fano condition holds if and only
if
\[
\bigwedge_{\{i,j,k\}\in\mathcal{D}_{\mathrm{N}}}\mathrm{tri}(A_{i},A_{j},A_{k})\;\wedge\;A_{1}\perp A_{2}\perp A_{3}.
\]
\end{prop}
\begin{IEEEproof}
We only have to prove the ``if'' direction. Since $\mathrm{tri}(A_{i},A_{j},A_{k})$
ensures $A_{i},A_{j},A_{k}$ are uniform with the same cardinality,
we know that $A_{i}$ ($i\in\mathcal{E}$) are all uniform with the
same cardinality (let it be $q$). Since $A_{2}\stackrel{\iota}{\le}A_{1}A_{12}$
by $\mathrm{tri}(A_{1},A_{2},A_{12})$, and $A_{3}\stackrel{\iota}{\le}A_{1}A_{13}$
by $\mathrm{tri}(A_{1},A_{3},A_{13})$, we have $A_{1}A_{2}A_{3}\stackrel{\iota}{=}A_{1}A_{12}A_{13}$,
and hence there is a one-to-one correspondence between the tuple $(A_{1},A_{12},A_{13})$
and $(A_{1},A_{2},A_{3})$, and the tuple $(A_{1},A_{12},A_{13})$
is also uniformly distributed over a set of size $q^{3}$, implying
$A_{1}\perp A_{12}\perp A_{13}$. Similarly, we have
\begin{align*}
 & A_{1}A_{2}A_{3}\stackrel{\iota}{\le}A_{1}A_{2}A_{23}\stackrel{\iota}{\le}A_{1}A_{2}A_{123}\\
 & \;\stackrel{\iota}{\le}A_{1}A_{12}A_{123}\stackrel{\iota}{\le}A_{1}A_{12}A_{23}\stackrel{\iota}{\le}A_{1}A_{2}A_{3},
\end{align*}
hence all terms above contain the same information. For each term
above, the three random variables are independent. This covers all
cases in $\mathcal{I}_{\mathrm{F}}$ by symmetry.
\end{IEEEproof}
\medskip{}

The Fano-non-Fano condition plays a similar role as the permuting
frame of equivalences in \cite{herrmann1987frames,herrmann1995undecidability}.\footnote{Very loosely speaking, the $\alpha_{1}$ in \cite{herrmann1995undecidability}
corresponds to $A_{2}A_{3}$ in this paper, whereas $\varepsilon_{12}$
in \cite{herrmann1995undecidability} corresponds to $A_{12}A_{3}$
in this paper.} The rest of the proof of Theorem \ref{thm:undecide_ci}  is inspired
by the arguments in \cite{herrmann1995undecidability}, but with different
presentation and proofs. 

Given $A_{1},A_{2},A_{3},A_{12},A_{13},A_{23},A_{123}$ (denote this
collection as $\{A_{i}\}_{i\in\mathcal{E}}$) satisfying the Fano-non-Fano
condition, we call $(\mathcal{A},\{\theta_{i}\}_{i\in\mathcal{E}})$
an \emph{abelian group labeling}, where $\mathcal{A}$ is an abelian
group, and $\theta_{i}$ is a bijective function mapping values of
$A_{i}$ to $\mathcal{A}$, if we have
\begin{align}
\theta_{12}(A_{12}) & =\theta_{1}(A_{1})+\theta_{2}(A_{2}),\nonumber \\
\theta_{13}(A_{13}) & =\theta_{1}(A_{1})+\theta_{3}(A_{3}),\nonumber \\
\theta_{23}(A_{23}) & =\theta_{2}(A_{2})+\theta_{3}(A_{3}),\nonumber \\
\theta_{123}(A_{123}) & =\theta_{1}(A_{1})+\theta_{2}(A_{2})+\theta_{3}(A_{3})\label{eq:abelian_label}
\end{align}
with probability $1$. The goal of this section is to show the following
proposition.
\begin{prop}
\label{prop:labeling}If $\{A_{i}\}_{i\in\mathcal{E}}$ satisfy the
Fano-non-Fano condition, then there exists an abelian group labeling
$(\mathcal{A},\{\theta_{i}\}_{i\in\mathcal{E}})$.
\end{prop}
An equivalent form of \eqref{eq:abelian_label} has appeared in \cite[Def. 6]{dougherty2006unachievability}.
The proof of Proposition \ref{prop:labeling} also shares a number
of similarities with \cite[Prop. 5]{dougherty2006unachievability}.
While it is possible to use the arguments in \cite[Prop. 5]{dougherty2006unachievability}
to prove Proposition \ref{prop:labeling} in this paper, we include
our proof of Proposition \ref{prop:labeling} for the sake of completeness.

 In the remainder of this section, we assume the Fano-non-Fano condition
holds, and construct an abelian group labeling. Since $\mathrm{tri}(A_{i},A_{j},A_{k})$
ensures $A_{i},A_{j},A_{k}$ are uniform with the same cardinality,
we know that $\{A_{i}\}_{i\in\mathcal{E}}$ are all uniform with the
same cardinality. For any $\{i,j,k\}\in\mathcal{D}_{\mathrm{N}}$,
since $\mathrm{tri}(A_{i},A_{j},A_{k})$ holds, $A_{k}$ is a function
of $A_{i},A_{j}$, and hence we can let this function be $f_{k}^{i,j}(a_{i},a_{j})$.
We use the notation $f_{k}^{i,j}$ for a function mapping values
of $(A_{i},A_{j})$ to values of $A_{k}$, where the superscript ``$i,j$''
denotes the ``domain'' and the subscript ``$k$'' denotes the
``codomain''. The choice of using superscript for domain and subscript
for codomain is due to the usual notation $\mathcal{B}^{\mathcal{A}}$
for the set of functions with domain $\mathcal{A}$ and codomain $\mathcal{B}$.

Note that there is a bijection between $(A_{1},A_{2},A_{3})$ and
$\{A_{i}\}_{i\in\mathcal{E}}$ since $A_{12}$ can be determined from
$A_{1},A_{2}$, and $A_{123}$ can be determined from $A_{12},A_{3}$.
Since $\{1,2,3\}\in\mathcal{I}_{\mathrm{F}}$, $A_{1}\perp A_{2}\perp A_{3}$,
and there are $k^{3}$ possible values of the tuple $(A_{1},A_{2},A_{3})$,
and hence there are $k^{3}$ possible values of the tuple $\{A_{i}\}_{i\in\mathcal{E}}$.
For any $i,j,k,l$ such that $\{i,j,k\}\in\mathcal{I}_{\mathrm{F}}$,
since there are $k^{3}$ possible values of the tuple $(A_{i},A_{j},A_{k})$
(the same as the number of tuples $\{A_{i}\}_{i\in\mathcal{E}}$),
there is a bijection between $(A_{i},A_{j},A_{k})$ and $\{A_{i}\}_{i\in\mathcal{E}}$,
we can let the function from $(A_{i},A_{j},A_{k})$ to $A_{l}$ be
$f_{l}^{i,j,k}(a_{i},a_{j},a_{k})$.

We first prove some properties of these functions.
\begin{prop}
\label{prop:f_prop}The following holds as long as all the functions
$f$ involved are defined:\footnote{Recall that $f_{k}^{i,j}$ is defined if and only if $\{i,j,k\}\in\mathcal{D}_{\mathrm{N}}$,
and $f_{l}^{i,j,k}$ is defined if and only if $\{i,j,k\}\in\mathcal{I}_{\mathrm{F}}$.}
\begin{enumerate}
\item $f_{k}^{i,j}(a,b)=f_{k}^{j,i}(b,a)$, and $f_{l}^{i,j,k}(a,b,c)=f_{l}^{j,k,i}(b,c,a)$
(similar for any permutation of arguments).
\item 
\begin{equation}
f_{i}^{k,j}(f_{k}^{i,j}(a,b),\,b)=a.\label{eq:f_flip}
\end{equation}
\item 
\begin{equation}
f_{l}^{i,j,k}(a,b,c)=f_{l}^{m,k}(f_{m}^{i,j}(a,b),c).\label{eq:f3}
\end{equation}
\end{enumerate}
\end{prop}
\begin{IEEEproof}
The first statement follows directly from the definition. For the
second statement, if $A_{i}=a$ and $A_{j}=b$ (note that this has
positive probability since $f_{k}^{i,j}$ is defined, meaning that
$\{i,j,k\}\in\mathcal{D}_{\mathrm{N}}$, which implies $A_{i}\perp A_{j}$)
implies $A_{k}=f_{k}^{i,j}(a,b)$, then $\mathbb{P}((A_{i},A_{j},A_{k})=(a,b,f_{k}^{i,j}(a,b)))>0$,
giving $f_{i}^{k,j}(f_{k}^{i,j}(a,b),\,b)=a$ since $f_{i}^{k,j}(A_{k},A_{j})=A_{i}$
must hold with probability $1$. For the third statement, if $A_{i}=a$,
$A_{j}=b$, $A_{k}=c$ (note that this has positive probability since
$f_{l}^{i,j,k}$ is defined, meaning that $\{i,j,k\}\in\mathcal{I}_{\mathrm{F}}$,
which implies $A_{i}\perp A_{j}\perp A_{k}$), then $A_{m}=f_{m}^{i,j}(a,b)$
and $A_{l}=f_{l}^{i,j,k}(a,b,c)$, giving $f_{l}^{m,k}(f_{m}^{i,j}(a,b),c)=f_{l}^{i,j,k}(a,b,c)$.
\end{IEEEproof}
\medskip{}

Since the labels of random variables do not matter in conditional
independence statements, we can assign any labels to the random variables.
We now assign labels to the random variables in $\{A_{i}\}_{i\in\mathcal{E}}$.
Our goal is to assign labels such that \eqref{eq:abelian_label} holds
without the functions $\theta$'s, i.e., $A_{12}=A_{1}+A_{2}$, $A_{13}=A_{1}+A_{3}$,
$A_{23}=A_{2}+A_{3}$, and $A_{123}=A_{1}+A_{2}+A_{3}$. Without loss
of generality, assume $0$ is an element in the supports of $A_{1}$,
$A_{2}$ and $A_{3}$. Let the support of $A_{123}$ be $\mathcal{A}$.
We label the value of $A_{123}$ when $A_{1}=A_{2}=A_{3}=0$ as $A_{123}=0$,
i.e.,
\begin{equation}
f_{123}^{1,2,3}(0,0,0)=0.\label{eq:f123_0}
\end{equation}
The other labels in $\mathcal{A}$ are arbitrary.

We now assign labels to the values of $A_{12}$ such that for any
$a\in\mathcal{A}$,
\begin{equation}
f_{12}^{123,3}(a,0)=a,\label{eq:a12_def}
\end{equation}
i.e., we label the value $f_{12}^{123,3}(a,0)$ of $A_{12}$ as $a$.
After this labeling, the support of $A_{12}$ is also $\mathcal{A}$.
This is a valid labeling since $a\mapsto f_{12}^{123,3}(a,0)$ is
a bijective function that maps values of $A_{123}$ to values of $A_{12}$
(it is bijective since the inverse of this function is $a\mapsto f_{123}^{12,3}(a,0)$
by \eqref{eq:f_flip}).  Similarly, we label $A_{13}$ and $A_{23}$
such that $f_{13}^{123,2}(a,0)=f_{23}^{123,1}(a,0)=a$. 

We then label $A_{1}$, $A_{2}$ and $A_{3}$ such that 
\begin{equation}
f_{1}^{123,23}(a,0)=f_{2}^{123,13}(a,0)=f_{3}^{123,12}(a,0)=a\label{eq:a1_def}
\end{equation}
 for $a\in\mathcal{A}$. Note that $0$ is in the support of $A_{23}$
since the support of $A_{23}$ is $\mathcal{A}$. While we have previously
fixed an element $0$ in the support of $A_{1}$, we do not change
the label of this element since 
\begin{align*}
f_{1}^{123,23}(0,0) & =f_{1}^{123,23}(0,f_{23}^{123,1}(0,0))=0
\end{align*}
 by \eqref{eq:f_flip}. 

The labeling is constructed such that the function $a\mapsto f_{k}^{i,j}(a,0)$
is the identity function for several triples $i,j,k$. We now show
that this holds for some more triples $i,j,k$.
\begin{prop}
\label{prop:fzero}For any distinct $i,j,k\in\{1,2,3\}$,
\begin{equation}
f_{ij}^{i,j}(a,0)=f_{ijk}^{ij,k}(a,0)=a,\label{eq:fijk_id}
\end{equation}
where the $ij$ in the subscripts and superscripts denote concatenation,
where the order of $i,j$ is ignored (e.g. $f_{ij}^{i,j}$ is $f_{12}^{2,1}$
when $i=2$, $j=1$).
\end{prop}
\begin{IEEEproof}
We have 
\begin{equation}
f_{123}^{12,3}(a,0)=f_{123}^{12,3}(f_{12}^{123,3}(a,0),0)=a\label{eq:f123_int}
\end{equation}
 by \eqref{eq:a12_def} and \eqref{eq:f_flip}. Also, 
\begin{align*}
f_{12}^{1,2}(a,0) & \stackrel{(a)}{=}f_{123}^{12,3}(f_{12}^{1,2}(a,0),0)\\
 & \stackrel{(b)}{=}f_{123}^{1,23}(a,f_{23}^{2,3}(0,0))\\
 & \stackrel{(a)}{=}f_{123}^{1,23}(a,f_{123}^{23,1}(f_{23}^{2,3}(0,0),0))\\
 & \stackrel{(b)}{=}f_{123}^{1,23}(a,f_{123}^{1,2,3}(0,0,0))\\
 & \stackrel{(c)}{=}f_{123}^{1,23}(a,0)\\
 & \stackrel{(d)}{=}f_{123}^{1,23}(f_{1}^{123,23}(a,0),0)\\
 & \stackrel{(e)}{=}a,
\end{align*}
where the lines marked by (a) are by \eqref{eq:f123_int}, (b) are
by \eqref{eq:f3}, (c) is by \eqref{eq:f123_0}, (d) is by \eqref{eq:a1_def},
and (e) is by \eqref{eq:f_flip}. The result follows from symmetry.
\end{IEEEproof}
\medskip{}

Moreover, it turns out that many of $f_{k}^{i,j}$ are actually the
same function under this labeling.
\begin{prop}
\label{prop:f_same}The following functions are the same:
\begin{align*}
 & f_{12}^{1,2}=f_{12}^{2,1}=f_{13}^{1,3}=f_{13}^{3,1}=f_{23}^{2,3}=f_{23}^{3,2}\\
 & =f_{123}^{1,23}=f_{123}^{2,13}=f_{123}^{3,12}=f_{123}^{23,1}=f_{123}^{13,2}=f_{123}^{12,3}.
\end{align*}
\end{prop}
\begin{IEEEproof}
Using Proposition \ref{prop:fzero},

\begin{align*}
f_{12}^{1,2}(a,b) & \stackrel{(a)}{=}f_{123}^{12,3}(f_{12}^{1,2}(a,b),0)\\
 & \stackrel{(b)}{=}f_{123}^{1,23}(a,f_{23}^{2,3}(b,0))\\
 & \stackrel{(a)}{=}f_{123}^{1,23}(a,b),
\end{align*}
where equalities marked with (a) are by Proposition \ref{prop:fzero},
and (b) is by \eqref{eq:f3}. Similarly $f_{13}^{1,3}(a,b)=f_{123}^{1,23}(a,b)$.
Hence $f_{12}^{1,2}(a,b)=f_{13}^{1,3}(a,b)$. By repeated use of this
fact and Proposition \ref{prop:f_prop}.1, we have $f_{12}^{1,2}(a,b)=f_{13}^{1,3}(a,b)=f_{13}^{3,1}(b,a)=f_{23}^{3,2}(b,a)=f_{23}^{2,3}(a,b)$.
By repeated use of this fact, we have $f_{12}^{1,2}=f_{13}^{1,3}=f_{23}^{2,3}=f_{12}^{2,1}=f_{13}^{3,1}=f_{23}^{3,2}$.
Also we have $f_{12}^{1,2}(a,b)=f_{12}^{2,1}(a,b)=f_{12}^{1,2}(b,a)=f_{123}^{1,23}(b,a)=f_{123}^{23,1}(a,b)$,
and hence $f_{12}^{1,2}=f_{123}^{1,23}=f_{123}^{23,1}$. The result
follows from repeated use of this fact.
\end{IEEEproof}
\medskip{}

We define an abelian group over $\mathcal{A}$ by 
\[
a+b:=f_{12}^{1,2}(a,b),
\]
\[
-a:=f_{2}^{1,12}(a,0),
\]
and the identity element is $0$.  A consequence of Proposition \ref{prop:f_prop}.1
and Proposition \ref{prop:f_same} is that $+$ is commutative, i.e.,
$a+b=b+a$. To check that $(\mathcal{A},+,0)$ is an abelian group,
we first prove $+$ is associative.
\begin{prop}
\label{prop:f_assoc}We have $(a+b)+c=a+(b+c)$, i.e.,
\[
f_{12}^{1,2}(f_{12}^{1,2}(a,b),c)=f_{12}^{1,2}(a,f_{12}^{1,2}(b,c)).
\]
\end{prop}
\begin{IEEEproof}
Using Proposition \ref{prop:f_same},
\begin{align*}
f_{12}^{1,2}(f_{12}^{1,2}(a,b),c) & =f_{123}^{12,3}(f_{12}^{1,2}(a,b),c)\\
 & \stackrel{(a)}{=}f_{123}^{1,23}(a,f_{23}^{2,3}(b,c))\\
 & =f_{12}^{1,2}(a,f_{12}^{1,2}(b,c)),
\end{align*}
where (a) is by \eqref{eq:f3}.
\end{IEEEproof}
\medskip{}

We then prove $-a$ is the additive inverse of $a$.
\begin{prop}
\label{prop:f_assoc-1}We have $a+(-a)=0$, i.e.,
\[
f_{12}^{1,2}(a,f_{2}^{1,12}(a,0))=0.
\]
\end{prop}
\begin{IEEEproof}
Using Proposition \ref{prop:f_same},
\begin{align*}
f_{12}^{1,2}(a,f_{2}^{1,12}(a,0)) & =f_{123}^{13,2}(a,f_{2}^{1,12}(a,0))\\
 & \stackrel{(a)}{=}f_{123}^{3,12}(f_{3}^{13,1}(a,a),0)\\
 & =f_{3}^{13,1}(a,a)\\
 & =f_{3}^{13,1}(f_{13}^{1,3}(a,0),a)\\
 & \stackrel{(b)}{=}0,
\end{align*}
where (a) is by \eqref{eq:f3}, and (b) is by \eqref{eq:f_flip}.
\end{IEEEproof}
Hence we conclude that $(\mathcal{A},+,0)$ is indeed an abelian group.
In this abelian group, we have $A_{12}=A_{1}+A_{2}$, $A_{13}=A_{1}+A_{3}$,
$A_{23}=A_{2}+A_{3}$, and $A_{123}=A_{1}+A_{2}+A_{3}$. This concludes
the proof of Proposition \ref{prop:labeling}.

We then show that the converse of Proposition \ref{prop:labeling}
holds as well, that is, any finite abelian group can be represented
by random variables satisfying the Fano-non-Fano condition \eqref{eq:fnf}.
\begin{prop}
\label{prop:abelian_to_fnf}For any finite abelian group $\mathcal{A}$,
there exist $\{A_{i}\}_{i\in\mathcal{E}}$ satisfying the Fano-non-Fano
condition and an abelian group labeling $(\mathcal{A},\{\theta_{i}\}_{i\in\mathcal{E}})$.
\end{prop}
\begin{IEEEproof}
Take $A_{1},A_{2},A_{3}$ to be i.i.d. uniformly drawn elements in
$\mathcal{A}$, and $A_{12}=A_{1}+A_{2}$, $A_{13}=A_{1}+A_{3}$,
$A_{23}=A_{2}+A_{3}$, $A_{123}=A_{1}+A_{2}+A_{3}$. We have $\mathrm{tri}(A_{1},A_{2},A_{12})$
since $A_{1}=A_{12}-A_{2}$, $A_{2}=A_{12}-A_{1}$. Similarly $\mathrm{tri}(A_{1},A_{23},A_{123})$
holds. By Proposition \ref{prop:fnf_equiv}, the Fano-non-Fano condition
is satisfied, and $(\mathcal{A},\{\theta_{i}\}_{i\in\mathcal{E}})$,
with $\theta_{i}$ being the identity function, is an abelian group
labeling.
\end{IEEEproof}
\[
\]

\section{Reduction from the Word Problem for Finite Monoids\label{sec:monoid}}

\subsection{The Word Problem and the Endomorphism Monoid}

This section follows the same high-level ideas as \cite[Section 6]{herrmann1995undecidability}
(i.e., reduction from the uniform word problem for finite monoids,
and the use of the endomorphism monoid), but with different arguments
concerning random variables (instead of equivalence relations in \cite{herrmann1995undecidability}).

The uniform word problem for groups/semigroups/monoids \cite{markov1951impossibility,post1947recursive,turing1950word}
is to determine whether two words represent the same element in the
group/semigroup/monoid, given a presentation of the group/semigroup/monoid.
In this paper, we will utilize the uniform word problem for finite
monoids in \cite{gurevich1966problem}. Recall that a finite monoid
$(\mathcal{M},\cdot,e)$ is a finite set $\mathcal{M}$ with an associative
binary operation ``$\cdot$'' (i.e., $a\cdot(b\cdot c)=(a\cdot b)\cdot c$)
and an identity element $e\in\mathcal{M}$ (i.e., $a\cdot e=e\cdot a=a$).
A sequence of elements $x_{1}\cdots x_{k}$ is called a word, and
it corresponds to the product $\prod_{i=1}^{k}x_{i}=x_{1}\cdot\cdots\cdot x_{k}$
in the monoid. The word problem is to decide, given a list of equalities
between the product represented by words, whether another such equality
follows from the list. More formally, given $l,k\in\mathbb{Z}_{>0}$,
$a_{i,j}\in\{1,\ldots,k\}$ for $i=0,\ldots,l$ and $j=1,\ldots,m_{i}$,
and $b_{i,j}\in\{1,\ldots,k\}$ for $i=0,\ldots,l$ and $j=1,\ldots,n_{i}$,
the uniform word problem for finite monoids is to determine whether
the implication
\begin{equation}
\bigwedge_{i=1}^{l}\left(\prod_{j=1}^{m_{i}}x_{a_{i,j}}=\prod_{j=1}^{n_{i}}x_{b_{i,j}}\right)\;\to\;\prod_{j=1}^{m_{i}}x_{a_{0,j}}=\prod_{j=1}^{n_{i}}x_{b_{0,j}}\label{eq:impl}
\end{equation}
holds for all finite monoid $(\mathcal{M},\cdot,e)$ and all $k$-tuples
$x_{1},\ldots,x_{k}\in\mathcal{M}$. It was shown in \cite{gurevich1966problem}
that the uniform word problem for finite monoids is undecidable, i.e.,
there does not exist an algorithm that, given $l,k,\{a_{i,j}\},\{b_{i,j}\}$
as input, outputs whether the implication holds for all $(\mathcal{M},\cdot,e)$
and $x_{1},\ldots,x_{k}\in\mathcal{M}$.

Note that we can restate the uniform word problem in the following
equivalent form: Given $l,k\in\mathbb{Z}_{>0}$, $a_{i},b_{i},c_{i}\in\{1,\ldots,k\}$
for $i=0,\ldots,l$, determine whether the implication
\begin{equation}
\bigwedge_{i=1}^{l}\left(x_{a_{i}}\cdot x_{b_{i}}=x_{c_{i}}\right)\;\to\;x_{a_{0}}=x_{c_{0}}\label{eq:impl_sim}
\end{equation}
holds for all finite monoid $(\mathcal{M},\cdot,e)$ and all $k$-tuples
$x_{1},\ldots,x_{k}\in\mathcal{M}$. To show how we can translate
\eqref{eq:impl} into \eqref{eq:impl_sim}, define intermediate variables
$y_{i,j}$ for $i=0,\ldots,l$ and $j=1,\ldots,m_{i}$ subject to
the constraints $y_{i,1}=x_{a_{i,1}}$ and $y_{i,j}=y_{i,j-1}\cdot x_{a_{i,j}}$
for $j=2,\ldots,m_{i}$. Then we can replace $\prod_{j=1}^{m_{i}}x_{a_{i,j}}$
by $y_{i,m_{i}}$. Similarly define $z_{i,j}$ for $i=0,\ldots,l$
and $j=1,\ldots,n_{i}$ and replace $\prod_{j=1}^{n_{i}}x_{b_{i,j}}$
by $z_{i,n_{i}}$. Finally, for any equality constraint (e.g. $y_{i,m_{i}}=z_{i,n_{i}}$),
replace all occurrences of one of them by the other.

We will show the desired undecidability result via a reduction from
the uniform word problem for finite monoids. In Section \ref{sec:fano},
we have shown that random variables satisfying the Fano-non-Fano condition
correspond to a finite abelian group. We then consider the endomorphism
monoid of an abelian group. Recall that a homomorphism $g:\mathcal{A}\to\mathcal{B}$,
where $\mathcal{A},\mathcal{B}$ are abelian groups, is a function
satisfying $g(a+b)=g(a)+g(b)$. An endomorphism in $\mathcal{A}$
is a homomorphism $g:\mathcal{A}\to\mathcal{A}$. The \emph{endomorphism
monoid} of an abelian group $\mathcal{A}$, denoted as $\mathrm{End}(\mathcal{A})$,
is the set of endomorphisms in $\mathcal{A}$, equipped with the operation
$g\cdot h:\mathcal{A}\to\mathcal{A}$ where $g\cdot h(a)=g(h(a))$. 

The following proposition in \cite{kurosh1963lectures} and \cite[Prop. 19]{herrmann1995undecidability}
shows that there is no loss of generality in considering endomorphism
monoids instead of general finite monoids, in the sense that any finite
monoid can be embedded into the endomorphism monoid of a finite abelian
group. Refer to \cite[Prop. 19]{herrmann1995undecidability} for the
proof. Recall that an embedding from a monoid to another monoid is
an injective function $h:\mathcal{M}\to\mathcal{N}$ satisfying $h(a\cdot b)=h(a)\cdot h(b)$
and $h(e_{\mathcal{M}})=e_{\mathcal{N}}$, where $e_{\mathcal{M}}$
is the identity element in $\mathcal{M}$.
\begin{prop}
[\cite{kurosh1963lectures}] For any finite monoid, there exists
an embedding from that monoid into the endomorphism monoid of a finite
abelian group.
\end{prop}
\medskip{}

Therefore, the implication \eqref{eq:impl_sim} holds for all finite
monoids if and only if \eqref{eq:impl_sim} holds for all finite abelian
group $\mathcal{A}$ and all $k$-tuples $x_{1},\ldots,x_{k}\in\mathrm{End}(\mathcal{A})$.
To show this, note that if \eqref{eq:impl_sim} holds for all finite
monoids, then it clearly holds for all endomorphism monoids of finite
abelian groups. If \eqref{eq:impl_sim} holds for all endomorphism
monoids of finite abelian groups, then for any finite monoid $\mathcal{M}$,
find an embedding $h:\mathcal{M}\to\mathrm{End}(\mathcal{A})$ where
$\mathcal{A}$ is a finite abelian group. For $x_{1},\ldots,x_{k}\in\mathcal{M}$,
if $\bigwedge_{i=1}^{l}(\prod_{j=1}^{m_{i}}x_{a_{i,j}}=\prod_{j=1}^{n_{i}}x_{b_{i,j}})$,
then $\bigwedge_{i=1}^{l}(\prod_{j=1}^{m_{i}}h(x_{a_{i,j}})=\prod_{j=1}^{n_{i}}h(x_{b_{i,j}}))$,
implying $\prod_{j=1}^{m_{i}}h(x_{a_{0,j}})=\prod_{j=1}^{n_{i}}h(x_{b_{0,j}})$,
which gives $h(\prod_{j=1}^{m_{i}}x_{a_{0,j}})=h(\prod_{j=1}^{n_{i}}x_{b_{0,j}})$,
and $\prod_{j=1}^{m_{i}}x_{a_{0,j}}=\prod_{j=1}^{n_{i}}x_{b_{0,j}}$
since $h$ is injective. Hence \eqref{eq:impl_sim} also holds for
all finite monoids.

\medskip{}

\subsection{Representing Endomorphisms as Random Variables}

The next step is to represent an endomorphism in $\mathcal{A}$ using
a random variable. If $(\mathcal{A},\{\theta_{i}\}_{i\in\mathcal{E}})$
is an abelian group labeling with $\theta_{i}$ being identity functions
(i.e., $A_{i}\in\mathcal{A}$), then we would represent an endomorphism
$g:\mathcal{A}\to\mathcal{A}$ by a random variable $U=A_{1}-g(A_{2})$.
To check whether $U$ corresponds to an endomorphism (up to relabeling)
using conditional independence relations, define the predicate $\mathrm{end}_{1,2}(\{A_{i}\}_{i\in\mathcal{E}},U)$
by \footnote{This step is inspired by \cite[Lemma 31]{herrmann1995undecidability}.}
\begin{align}
 & \!\!\!\mathrm{end}_{1,2}(\{A_{i}\}_{i\in\mathcal{E}},U):\nonumber \\
 & \exists V,W:\,\mathrm{fnf}(\{A_{i}\}_{i\in\mathcal{E}})\,\nonumber \\
 & \;\wedge\,\mathrm{ueq}(U,A_{1})\,\wedge\,\mathrm{ueq}(V,A_{1})\,\wedge\,\mathrm{ueq}(W,A_{1})\nonumber \\
 & \;\wedge\,U\stackrel{\iota}{=}A_{1}|A_{2}\,\wedge\,V\stackrel{\iota}{=}A_{1}|A_{23}\,\wedge\,U\stackrel{\iota}{=}V|A_{3}\nonumber \\
 & \;\wedge\,W\stackrel{\iota}{=}A_{13}|A_{2}\,\wedge\,U\stackrel{\iota}{=}W|A_{3},\label{eq:end}
\end{align}
where we define $X\stackrel{\iota}{=}Y|Z\;\Leftrightarrow\;X\stackrel{\iota}{\le}ZY\,\wedge\,Y\stackrel{\iota}{\le}ZX$,
i.e., if we are given $Z$, then $X$ has the same information as
$Y$, and $\mathrm{ueq}(X,Y)$ is a predicate that checks whether
$X,Y$ are both uniform and have the same cardinality (this expression
was given in \cite{li2021undecidability}; refer to \cite{li2021undecidability}
for the proof):
\begin{align*}
 & \!\!\!\!\mathrm{ueq}(X,Y):\,\\
 & \exists U_{1},U_{2},U_{3}:\,\mathrm{tri}(X,U_{1},U_{2})\,\wedge\,\mathrm{tri}(Y,U_{1},U_{3}).
\end{align*}

We show that a random variable $U$ satisfying $\mathrm{end}_{1,2}(\{A_{i}\}_{i\in\mathcal{E}},U)$
corresponds to an endomorphism in $\mathcal{A}$.
\begin{prop}
\label{prop:end_label}Given $\{A_{i}\}_{i\in\mathcal{E}},U$ satisfying
$\mathrm{end}_{1,2}(\{A_{i}\}_{i\in\mathcal{E}},U)$, and an abelian
group labeling $(\mathcal{A},\{\theta_{i}\}_{i\in\mathcal{E}})$ of
$\{A_{i}\}_{i\in\mathcal{E}}$. Then there exists a unique endomorphism
$g:\mathcal{A}\to\mathcal{A}$, and a unique bijective function $\phi$
mapping the values of $U$ to $\mathcal{A}$, satisfying 
\begin{equation}
\phi(U)=\theta_{1}(A_{1})-g(\theta_{2}(A_{2}))\label{eq:end_unique}
\end{equation}
with probability $1$.
\end{prop}
\begin{IEEEproof}
Without loss of generality, assume $\{\theta_{i}\}_{i\in\mathcal{E}}$
are identity functions. Let the supports of $U,V,W$ be $\mathcal{U},\mathcal{V},\mathcal{W}$
respectively, with $|\mathcal{U}|=|\mathcal{V}|=|\mathcal{W}|=|\mathcal{A}|$
since $\mathrm{ueq}(U,A_{1})\,\wedge\,\mathrm{ueq}(V,A_{1})\,\wedge\,\mathrm{ueq}(W,A_{1})$.
Since $U\stackrel{\iota}{=}A_{1}|A_{2}$, we can find injective functions
$\mu_{a_{2}}:\mathcal{A}\to\mathcal{U}$ for $a_{2}\in\mathcal{A}$
such that $U=\mu_{A_{2}}(A_{1})$. Note that $\mu_{a_{2}}$ is bijective
since $|\mathcal{A}|=|\mathcal{U}|$.  Without loss of generality,
assume $\mu_{0}(a)=a$ for $a\in\mathcal{A}$, i.e., we assign the
label $a$ to the value $\mu_{0}(a)$ of $U$. After this labeling,
we have $\mathcal{U}=\mathcal{A}$. Similarly, we can assume $\mathcal{V}=\mathcal{A}$,
and define the bijective functions $\nu_{a_{23}}:\mathcal{A}\to\mathcal{A}$
such that $V=\nu_{A_{2}+A_{3}}(A_{1})$ and $\nu_{0}(a)=a$. Similarly,
we can assume $\mathcal{W}=\mathcal{A}$, and define the bijective
functions $\omega_{a_{2}}:\mathcal{A}\to\mathcal{A}$ such that $W=\omega_{A_{2}}(A_{1}+A_{3})$
and $\omega_{0}(a)=a$.

Consider $U\stackrel{\iota}{=}W|A_{3}$. Let $\tau:\mathcal{A}^{2}\to\mathcal{A}$
such that $W=\tau(A_{3},U)$. We have, for any $a_{1},a_{2},a_{3}\in\mathcal{A}$,
\[
\omega_{a_{2}}(a_{1}+a_{3})=\tau(a_{3},\,\mu_{a_{2}}(a_{1})).
\]
Substituting $a_{2}=0$, we have $a_{1}+a_{3}=\tau(a_{3},a_{1})$.
Hence,
\[
\omega_{a_{2}}(a_{1}+a_{3})=a_{3}+\mu_{a_{2}}(a_{1}).
\]
Substituting $a_{3}=-a_{1}$,
\[
\omega_{a_{2}}(0)=-a_{1}+\mu_{a_{2}}(a_{1}).
\]
Hence $\mu_{a_{2}}(a_{1})=a_{1}+\omega_{a_{2}}(0)$. Substituting
$a_{1}=0$, we have $\omega_{a_{2}}(0)=\mu_{a_{2}}(0)$. Hence,
\[
\mu_{a_{2}}(a_{1})=a_{1}+\mu_{a_{2}}(0).
\]

Consider $U\stackrel{\iota}{=}V|A_{3}$. Let $\kappa:\mathcal{A}^{2}\to\mathcal{A}$
such that $V=\kappa(A_{3},U)$. We have, for any $a_{1},a_{2},a_{3}\in\mathcal{A}$,
\begin{align*}
\nu_{a_{2}+a_{3}}(a_{1}) & =\kappa(a_{3},\,\mu_{a_{2}}(a_{1}))\\
 & =\kappa(a_{3},\,a_{1}+\mu_{a_{2}}(0)).
\end{align*}
Substituting $a_{2}=0$, we have $\nu_{a_{3}}(a_{1})=\kappa(a_{3},a_{1})$.
Hence,
\[
\nu_{a_{2}+a_{3}}(a_{1})=\nu_{a_{3}}(a_{1}+\mu_{a_{2}}(0)).
\]
Substituting $a_{3}=0$, we have $\nu_{a_{2}}(a_{1})=a_{1}+\mu_{a_{2}}(0)$.
Therefore,
\[
a_{1}+\mu_{a_{2}+a_{3}}(0)=a_{1}+\mu_{a_{2}}(0)+\mu_{a_{3}}(0),
\]
and hence $\mu_{a+b}(0)=\mu_{a}(0)+\mu_{b}(0)$. As a result, if $\mathrm{end}_{1,2}(\{A_{i}\}_{i\in\mathcal{E}},U)$
is satisfied, and $(\mathcal{A},\{\theta_{i}\}_{i\in\mathcal{E}})$
(where $\{\theta_{i}\}_{i\in\mathcal{E}}$ are identity functions)
is an abelian group labeling of $\{A_{i}\}_{i\in\mathcal{E}}$, then
we can find an endomorphism $g(a)=-\mu_{a}(0)$ in $\mathcal{A}$
satisfying $U=\mu_{A_{2}}(A_{1})=A_{1}-g(A_{2})$ up to relabeling.

For uniqueness, assume $\phi'$ and $g'$ satisfy \eqref{eq:end_unique},
i.e., $\phi'(U)=A_{1}-g'(A_{2})$. We have
\begin{equation}
\phi'(a_{1}-g(a_{2}))=a_{1}-g'(a_{2}).\label{eq:phiprime}
\end{equation}
Substituting $a_{2}=0$, since $g,g'$ are endomorphisms, we have
$\phi'(a_{1})=a_{1}$. Substituting back to \eqref{eq:phiprime},
\[
a_{1}-g(a_{2})=a_{1}-g'(a_{2}),
\]
\[
g'(a_{2})=g(a_{2}).
\]
Hence the choice of $\phi,g$ is unique. 
\end{IEEEproof}
We then show that the converse of Proposition \ref{prop:end_label}
holds as well. Therefore there is a one-to-one correspondence (up
to relabelling) between endomorphisms and random variables satisfying
$\mathrm{end}_{1,2}(\{A_{i}\}_{i\in\mathcal{E}},U)$.
\begin{prop}
\label{prop:end_to_fnf}Given $\{A_{i}\}_{i\in\mathcal{E}}$ satisfying
$\mathrm{fnf}(\{A_{i}\}_{i\in\mathcal{E}})$, an abelian group labeling
$(\mathcal{A},\{\theta_{i}\}_{i\in\mathcal{E}})$, and an endomorphism
$g:\mathcal{A}\to\mathcal{A}$. Let $U=\theta_{1}(A_{1})-g(\theta_{2}(A_{2}))$.
Then $\mathrm{end}_{1,2}(\{A_{i}\}_{i\in\mathcal{E}},U)$ holds.
\end{prop}
\begin{IEEEproof}
Without loss of generality, assume $\{\theta_{i}\}_{i\in\mathcal{E}}$
are identity functions, and hence $U=A_{1}-g(A_{2})$. Take $V=A_{1}-g(A_{2}+A_{3})=A_{1}-g(A_{2})-g(A_{3})$,
and $W=A_{1}-g(A_{2})+A_{3}$. It is straightforward to check that
the conditions in \eqref{eq:end} hold.
\end{IEEEproof}
\medskip{}

Similar to \eqref{eq:end}, we can define the predicates $\mathrm{end}_{i,j}(\{A_{i}\},U)$
for $i\neq j\in\{1,2,3\}$, which checks whether there exists an endomorphism
$g:\mathcal{A}\to\mathcal{A}$ and a bijective function $\phi$ satisfying
$\phi(U)=\theta_{i}(A_{i})-g(\theta_{j}(A_{j}))$.

It is left to represent the composition of endomorphisms using random
variables.\footnote{This step is inspired by \cite[Lemma 21]{herrmann1995undecidability}.}

\begin{prop}
\label{prop:comp_123}Given $\{A_{i}\}_{i\in\mathcal{E}},U_{1},U_{2},U_{3}$
satisfying $\mathrm{end}_{1,2}(\{A_{i}\},U_{1})$, $\mathrm{end}_{2,3}(\{A_{i}\},U_{2})$,
$\mathrm{end}_{1,3}(\{A_{i}\},U_{3})$, and an abelian group labeling
$(\mathcal{A},\{\theta_{i}\}_{i\in\mathcal{E}})$. We have $U_{3}\stackrel{\iota}{\le}U_{1}U_{2}$
if and only if $g_{3}=g_{1}\cdot g_{2}$, where $g_{i}$ is the endomorphism
corresponding to $U_{i}$.
\end{prop}
\begin{IEEEproof}
Without loss of generality, assume $\{\theta_{i}\}_{i\in\mathcal{E}}$
and the mappings $\phi$ in Proposition \ref{prop:end_label} are
identity functions. Hence $U_{1}=A_{1}-g_{1}(A_{2})$, $U_{2}=A_{2}-g_{2}(A_{3})$,
$U_{3}=A_{1}-g_{3}(A_{3})$. For the ``if'' direction, if $g_{3}=g_{1}\cdot g_{2}$,
then $U_{3}=A_{1}-g_{1}(g_{2}(A_{3}))=U_{1}+g_{1}(U_{2})$ is a function
of $(U_{1},U_{2})$. 

For the ``only if'' direction, assume $U_{3}\stackrel{\iota}{\le}U_{1}U_{2}$.
Let $h$ be a function such that $U_{3}=h(U_{1},U_{2})$. We have
\begin{equation}
a_{1}-g_{3}(a_{3})=h\left(a_{1}-g_{1}(a_{2}),\,a_{2}-g_{2}(a_{3})\right).\label{eq:agh}
\end{equation}
Substituting $a_{1}=x+g_{1}(a_{2})$ and $a_{3}=0$,
\[
x+g_{1}(a_{2})=h\left(x,\,a_{2}\right).
\]
Substituting back to \eqref{eq:agh},
\begin{align*}
a_{1}-g_{3}(a_{3}) & =a_{1}-g_{1}(a_{2})+g_{1}(a_{2}-g_{2}(a_{3}))\\
 & =a_{1}-g_{1}(g_{2}(a_{3})).
\end{align*}
Therefore, we have $g_{3}=g_{1}\cdot g_{2}$.
\end{IEEEproof}
\medskip{}

The problem of Proposition \ref{prop:comp_123} is that $U_{1},U_{2},U_{3}$
are subject to $\mathrm{end}_{i,j}$ for different $i,j$. We want
a predicate that checks for composition of endomorphisms using $\mathrm{end}_{1,2}$
only. Therefore, we require a way to convert between $\mathrm{end}_{i,j}$
for different $i,j$. Define
\begin{align}
 & \!\!\mathrm{conv}_{1,3}^{1,2}(\{A_{i}\},U,V):\nonumber \\
 & \exists W:\,\mathrm{end}_{1,2}(\{A_{i}\},U)\,\wedge\,\mathrm{end}_{1,3}(\{A_{i}\},V)\,\wedge\,\mathrm{end}_{2,3}(\{A_{i}\},W)\nonumber \\
 & \wedge\,A_{13}\stackrel{\iota}{\le}A_{12}W\,\wedge\,V\stackrel{\iota}{\le}UW.\label{eq:conv1213}
\end{align}

\begin{prop}
\label{prop:swap_j}Given $\{A_{i}\}_{i\in\mathcal{E}},U,V$ satisfying
$\mathrm{end}_{1,2}(\{A_{i}\},U)$ and $\mathrm{end}_{1,3}(\{A_{i}\},V)$,
and an abelian group labeling $(\mathcal{A},\{\theta_{i}\}_{i\in\mathcal{E}})$.
We have $\mathrm{conv}_{1,3}^{1,2}(\{A_{i}\},U,V)$ if and only if
the endomorphism corresponding to $U,V$ are the same.
\end{prop}
\begin{IEEEproof}
Without loss of generality, assume $\{\theta_{i}\}_{i\in\mathcal{E}}$
and the mappings $\phi$ in Proposition \ref{prop:end_label} are
identity functions. Let $U=A_{1}-g(A_{2})$, $V=A_{1}-h(A_{3})$.
First show the ``only if'' direction. Note that $A_{12}$ satisfies
$\mathrm{end}_{1,2}(\{A_{i}\},A_{12})$ and corresponds to the negation
endomorphism $x\mapsto-x$, and $A_{13}$ satisfies $\mathrm{end}_{1,3}(\{A_{i}\},A_{12})$
and corresponds to the negation endomorphism $x\mapsto-x$. By Proposition
\eqref{prop:comp_123}, since $A_{13}\stackrel{\iota}{\le}A_{12}W$,
$W$ corresponds to the identity endomorphism (which, when composed
with the negation endomorphism, gives the negation endomorphism).
By Proposition \eqref{prop:comp_123}, since $V\stackrel{\iota}{\le}UW$,
$V$ corresponds to the same endomorphism as $U$. For the ``if''
direction, take $W$ such that $\mathrm{end}_{2,3}(\{A_{i}\},W)$
is satisfied, and it corresponds to the identity endomorphism.
\end{IEEEproof}
\medskip{}

Using $\mathrm{conv}_{1,3}^{1,2}$, we can convert between different
values for the second index $j$ in $\mathrm{end}_{i,j}$. To convert
between different first indices $i$, define
\begin{align}
 & \!\!\mathrm{conv}_{3,2}^{1,2}(\{A_{i}\},U,V):\nonumber \\
 & \exists W:\,\mathrm{end}_{1,2}(\{A_{i}\},U)\,\wedge\,\mathrm{end}_{3,2}(\{A_{i}\},V)\,\wedge\,\mathrm{end}_{1,3}(\{A_{i}\},W)\nonumber \\
 & \wedge\,A_{12}\stackrel{\iota}{\le}WA_{23}\,\wedge\,V\stackrel{\iota}{\le}WU.\label{eq:conv1232}
\end{align}

\begin{prop}
\label{prop:swap_i}Given $\{A_{i}\}_{i\in\mathcal{E}},U,V$ satisfying
$\mathrm{end}_{1,2}(\{A_{i}\},U)$ and $\mathrm{end}_{3,2}(\{A_{i}\},V)$,
and an abelian group labeling $(\mathcal{A},\{\theta_{i}\}_{i\in\mathcal{E}})$.
We have $\mathrm{conv}_{3,2}^{1,2}(\{A_{i}\},U,V)$ if and only if
the endomorphism corresponding to $U,V$ are the same.
\end{prop}
\begin{IEEEproof}
Without loss of generality, assume $\{\theta_{i}\}_{i\in\mathcal{E}}$
and the mappings $\phi$ in Proposition \ref{prop:end_label} are
identity functions. Let $U=A_{1}-g(A_{2})$, $V=A_{3}-h(A_{2})$.
First show the ``only if'' direction. As in the proof of Proposition
\ref{prop:swap_j}, we can assume $W=A_{1}-A_{3}$. Since $V\stackrel{\iota}{\le}WU$,
we can let $\kappa$ be a function such that $V=\kappa(W,U)$. We
have
\begin{equation}
a_{3}-h(a_{2})=\kappa\left(a_{1}-a_{3},\,a_{1}-g(a_{2})\right).\label{eq:agh-1}
\end{equation}
Substituting $a_{3}=a_{1}-x$ and $a_{2}=0$,
\[
a_{1}-x=\kappa\left(x,\,a_{1}\right).
\]
Substituting back to \eqref{eq:agh-1},
\begin{align*}
a_{3}-h(a_{2}) & =a_{1}-g(a_{2})-(a_{1}-a_{3}),
\end{align*}
giving $h=g$. For the ``if'' direction, take $W$ such that $\mathrm{end}_{1,3}(\{A_{i}\},W)$
is satisfied, and it corresponds to the identity endomorphism.
\end{IEEEproof}
\medskip{}

Combining these constructions, we can use the following predicate
to check whether $U_{1},U_{2},U_{3}$ with $\mathrm{end}_{1,2}(\{A_{i}\},U_{j})$
for $j=1,2,3$ satisfy $g_{3}=g_{1}\cdot g_{2}$, where $g_{i}$ is
the endomorphism corresponding to $U_{i}$:
\begin{align}
 & \!\!\mathrm{comp}_{1,2}(\{A_{i}\},U_{1},U_{2},U_{3}):\nonumber \\
 & \exists V_{1},V_{2}:\,\bigwedge_{j=1}^{3}\mathrm{end}_{1,2}(\{A_{i}\},U_{j})\,\nonumber \\
 & \wedge\,\mathrm{conv}_{1,3}^{1,2}(\{A_{i}\},U_{1},V_{1})\,\wedge\,\mathrm{conv}_{3,2}^{1,2}(\{A_{i}\},U_{2},V_{2})\nonumber \\
 & \wedge\,U_{3}\stackrel{\iota}{\le}V_{1}V_{2}.\label{eq:comp}
\end{align}
The following is a direct consequence of Propositions \ref{prop:comp_123},
\ref{prop:swap_j} and \ref{prop:swap_i}.
\begin{prop}
\label{prop:comp_12}Given $\{A_{i}\}_{i\in\mathcal{E}},U_{1},U_{2},U_{3}$
satisfying $\mathrm{end}_{1,2}(\{A_{i}\},U_{j})$ for $j=1,2,3$,
and an abelian group labeling $(\mathcal{A},\{\theta_{i}\}_{i\in\mathcal{E}})$.
We have $\mathrm{comp}_{1,2}(\{A_{i}\},U_{1},U_{2},U_{3})$ if and
only if $g_{3}=g_{1}\cdot g_{2}$, where $g_{i}$ is the endomorphism
corresponding to $U_{i}$.
\end{prop}
We then show how to check for equality between two endomorphisms with
the same $i,j$ in $\mathrm{end}_{i,j}$.
\begin{prop}
\label{prop:12eq}Given $\{A_{i}\}_{i\in\mathcal{E}},U,V$ satisfying
$\mathrm{end}_{1,2}(\{A_{i}\},U)$ and $\mathrm{end}_{1,2}(\{A_{i}\},V)$,
and an abelian group labeling $(\mathcal{A},\{\theta_{i}\}_{i\in\mathcal{E}})$.
We have $U\stackrel{\iota}{\le}V$ if and only if the endomorphism
corresponding to $U,V$ are the same.
\end{prop}
\begin{IEEEproof}
Without loss of generality, assume $\{\theta_{i}\}_{i\in\mathcal{E}}$
and the mappings $\phi$ in Proposition \ref{prop:end_label} are
identity functions. Let $U=A_{1}-g(A_{2})$, $V=A_{1}-h(A_{2})$.
The ``if'' direction follows directly from the uniqueness in Proposition
\ref{prop:end_label}. For the ``only if'' direction, assume $U\stackrel{\iota}{\le}V$,
and let $\kappa$ be a function such that $U=\kappa(V)$. We have
$a_{1}-g(a_{2})=\kappa(a_{1}-h(a_{2}))$. Substituting $a_{2}=0$,
we have $\kappa(a_{1})=a_{1}$. Hence $a_{1}-g(a_{2})=a_{1}-h(a_{2})$,
which gives $g=h$.
\end{IEEEproof}
\medskip{}

Combining \eqref{eq:comp} and Proposition \ref{prop:12eq}, we know
that \eqref{eq:impl_sim} holds for all finite abelian group $\mathcal{A}$
and all $x_{1},\ldots,x_{k}\in\mathrm{End}(\mathcal{A})$ if and only
if the implication
\begin{align}
 & \bigwedge_{j=1}^{k}\mathrm{end}_{1,2}(\{A_{i}\},U_{j})\nonumber \\
 & \wedge\,\bigwedge_{j=1}^{l}\mathrm{comp}_{1,2}(\{A_{i}\},U_{a_{j}},U_{b_{j}},U_{c_{j}})\nonumber \\
 & \to\;U_{a_{0}}\stackrel{\iota}{\le}U_{c_{0}}\label{eq:impl_final}
\end{align}
holds for all finite random variables $\{A_{i}\}$, $U_{1},\ldots,U_{k}$.
The complete proof of this equivalence is given below for the sake
of completeness.
\begin{prop}
The implication \eqref{eq:impl_sim} holds for all finite abelian
group $\mathcal{A}$ and all $x_{1},\ldots,x_{k}\in\mathrm{End}(\mathcal{A})$
if and only if the implication \eqref{eq:impl_final} holds for all
finite random variables $\{A_{i}\}$, $U_{1},\ldots,U_{k}$. 
\end{prop}
\begin{IEEEproof}
For the ``if'' direction, assume \eqref{eq:impl_final} holds for
all random variables, then for any finite abelian group $\mathcal{A}$
and $x_{1},\ldots,x_{k}\in\mathrm{End}(\mathcal{A})$ satisfying the
left hand side of \eqref{eq:impl_sim}, let $\{A_{i}\}$ satisfy $\mathrm{fnf}(\{A_{i}\})$
such that $(\mathcal{A},\{\theta_{i}\}_{i\in\mathcal{E}})$ is an
abelian group labeling (by Proposition \ref{prop:abelian_to_fnf}).
Let $U_{j}$ satisfy $\mathrm{end}_{1,2}(\{A_{i}\},U_{j})$ corresponding
to the endomorphism $x_{j}$ (by Proposition \ref{prop:end_to_fnf}).
Since $x_{a_{j}}\cdot x_{b_{j}}=x_{c_{j}}$, $\mathrm{comp}_{1,2}(\{A_{i}\},U_{a_{j}},U_{b_{j}},U_{c_{j}})$
holds by Proposition \ref{prop:comp_12}, the left hand side of \eqref{eq:impl_final}
holds, and hence $U_{a_{0}}\stackrel{\iota}{\le}U_{c_{0}}$, and $x_{a_{0}}=x_{c_{0}}$
by Proposition \ref{prop:12eq}.

For the ``only if'' direction, assume \eqref{eq:impl_sim} holds
for all finite abelian group $\mathcal{A}$ and all $x_{1},\ldots,x_{k}\in\mathrm{End}(\mathcal{A})$.
Fix any $\{A_{i}\}$, $U_{1},\ldots,U_{k}$ satisfying the left hand
side of \eqref{eq:impl_final}. Fix any abelian group labeling $(\mathcal{A},\{\theta_{i}\}_{i\in\mathcal{E}})$.
By Proposition \ref{prop:end_label}, let $x_{j}$ be the endomorphism
corresponding to $U_{j}$. Since $\mathrm{comp}_{1,2}(\{A_{i}\},U_{a_{j}},U_{b_{j}},U_{c_{j}})$,
we have $x_{a_{j}}\cdot x_{b_{j}}=x_{c_{j}}$ by Proposition \ref{prop:comp_12},
and hence the left hand side of \eqref{eq:impl_sim} holds, implying
$x_{a_{0}}=x_{c_{0}}$, which gives $U_{a_{0}}\stackrel{\iota}{\le}U_{c_{0}}$
by Proposition \ref{prop:12eq}.
\end{IEEEproof}
\medskip{}

We have shown a reduction from the word problem for endomorphism monoids
of abelian groups to the conditional independence implication problem,
which gives the desired undecidability result. Note that the actual
random variables in the conditional independence implication problem
(stated in the form in Theorem \ref{thm:undecide_ci}) is considerably
more than just $\{A_{i}\}$, $U_{1},\ldots,U_{k}$, since there are
many existentially-quantified intermediate random variables (e.g.
$V,W$ in \eqref{eq:end}) in the construction (existential quantification
becomes universal quantification since all these predicates appear
on the left hand side of the implication in \eqref{eq:impl_final}). 

Also note that Theorem \ref{thm:undecide_ci} requires $\mathcal{U}_{i},\mathcal{V}_{i},\mathcal{W}_{i}$
to be disjoint, though our construction involves $Y\stackrel{\iota}{\le}X\;\Leftrightarrow\;Y\perp Y|X$,
which is a non-disjoint conditional independence condition. Disjointness
is not an obstacle, since it was shown in \cite[Thm 4]{li2021undecidabilityarxiv}
that the conditional independence implication problem for disjoint
$\mathcal{U}_{i},\mathcal{V}_{i},\mathcal{W}_{i}$ can be reduced
from the conditional independence implication problem for general
(not necessarily disjoint) $\mathcal{U}_{i},\mathcal{V}_{i},\mathcal{W}_{i}$.
Also see \cite{herrmann2006corrigendum} for a related argument.

This completes the proof of Theorem \ref{thm:undecide_ci}.

\medskip{}

\section{Network Coding\label{sec:network}}

We use the same definition of network as \cite{dougherty2007networks}
with minor notational differences. A network is a directed acyclic
multigraph $(\mathcal{N},\mathcal{L})$, where $\mathcal{N}$ is the
vertex set and $\mathcal{L}$ is the edge set. Let $M_{i}$ ($i=1,\ldots,n$)
be the source messages, which are independent uniformly distributed
random variables with cardinality $q$, where $q\in\mathbb{Z}_{\ge2}$
is the alphabet size. Each node $v\in\mathcal{N}$ has access to a
subset of the messages with indices in the set $\mathcal{C}_{v}\subseteq\{1,\ldots,n\}$,
and demands another subset of the messages $\mathcal{D}_{v}\subseteq\{1,\ldots,n\}$.
In topological order of the multigraph, each node transmits a signal,
which is an element in a set of size $q$, along each outgoing edge,
where the signals can depend on the messages that the node has access
to, and the signals along incoming edges to the node. At the end of
the transmission, each node must decode the set of messages it demands.

More precisely, we let $X_{f}$ be the signal along edge $f\in\mathcal{L}$,
which is a random variable with cardinality at most $q$. The \emph{coding
constraint} is that
\begin{align}
\big(M_{\mathcal{D}_{v}},\,X_{\mathrm{out}(v)}\big) & \stackrel{\iota}{\le}\big(M_{\mathcal{C}_{v}},\,X_{\mathrm{in}(v)}\big)\label{eq:coding_cons}
\end{align}
for all $v\in\mathcal{N}$, where $\mathrm{in}(v)\subseteq\mathcal{L}$
is the set of incoming edges to the node $v$, and $\mathrm{out}(v)$
is the set of outgoing edges, and we write $X_{\mathrm{out}(v)}=\{X_{f}\}_{f\in\mathrm{out}(v)}$.
The problem is to decide whether the network is \emph{solvable}, that
is, whether there exists $q\in\mathbb{Z}_{\ge2}$, $M_{1},\ldots,M_{n}$
which are independent uniformly distributed random variables with
cardinality $q$, and $\{X_{f}\}_{f\in\mathcal{L}}$ with cardinalities
at most $q$, satisfying the coding constraint \eqref{eq:coding_cons}.

In this section, we will show that network coding is undecidable.
\begin{thm}
\label{thm:netcode}The following problem is undecidable: Given a
network, decide whether it is solvable.
\end{thm}
\medskip{}

We may also be interested in the case where the messages and signals
are sequences of symbols in an alphabet of size $c\in\mathbb{Z}_{\ge2}$
with the same length (e.g. $c=2$ if bit sequences are being sent).
Equivalently, we may consider a sequence to be an element in the overall
alphabet of size $q=c^{m}$ for some $m\in\mathbb{Z}_{\ge1}$. This
case is undecidable as well.
\begin{thm}
\label{thm:netcode_vec}For any fixed $c\in\mathbb{Z}_{\ge2}$, the
following problem is undecidable: Given a network, decide whether
there exists $m\in\mathbb{Z}_{\ge1}$ such that the network is solvable
with alphabet size $q=c^{m}$.
\end{thm}
\medskip{}

The proof is divided into the following subsections.

\subsection{The Left Regular Representation of a Finite Group}

Fix a finite field $\mathbb{F}$. For a finite set $\mathcal{S}$,
write $\mathbb{F}^{\mathcal{S}}=\{\{x_{a}\}_{a\in\mathcal{S}}:\,x_{a}\in\mathbb{F}\}$
for the $|\mathcal{S}|$-dimensional vector space where each element
$\mathbf{x}=\{x_{a}\}_{a\in\mathcal{S}}$ is a vector with entries
indexed by elements in $\mathcal{S}$. Given any finite group $\mathcal{B}$,
we consider the \emph{left regular representation} \cite{fulton2013representation},
a basic construction in representation theory, which is an embedding
from $\mathcal{B}$ into the general linear group $\mathrm{GL}(\mathbb{F}^{\mathcal{B}})$,
where $\mathrm{GL}(\mathbb{F}^{\mathcal{B}})$ consists of invertible
linear functions (or automorphisms) $\mathbb{F}^{\mathcal{B}}\to\mathbb{F}^{\mathcal{B}}$,
or equivalently, invertible $|\mathcal{B}|\times|\mathcal{B}|$ matrices,
with group operation given by function composition or matrix multiplication.
For any finite group $\mathcal{B}$, define an embedding $\lambda_{\mathcal{B}}:\mathcal{B}\to\mathrm{GL}(\mathbb{F}^{\mathcal{B}})$,
where $\lambda_{\mathcal{B}}(b)$ is the function $\{x_{a}\}_{a\in\mathcal{B}}\mapsto\{x_{b^{-1}\cdot a}\}_{a\in\mathcal{B}}$.
It is straightforward to check that $\lambda_{\mathcal{B}}$ is an
injective homomorphism.

We prove the following key observation about the left regular representation.
\begin{prop}
\label{prop:non_id}If $b\in\mathcal{B}$, $b\neq e_{\mathcal{B}}$
(the identity element of $\mathcal{B}$), then the function $\lambda_{\mathcal{B}}(b)-\mathrm{id}_{\mathbb{F}^{\mathcal{B}}}$
(i.e., the function $\mathbb{F}^{\mathcal{B}}\to\mathbb{F}^{\mathcal{B}}$,
$\mathbf{x}\mapsto\lambda_{\mathcal{B}}(b)(\mathbf{x})-\mathbf{x}$)
is a linear function with rank at least $|\mathcal{B}|/2$.
\end{prop}
\begin{IEEEproof}
Let $b\neq e_{\mathcal{B}}$. Consider the cyclic subgroup $\left\langle b\right\rangle =\{b^{n}:\,n\in\mathbb{Z}\}$
of $\mathcal{B}$. We have $|\left\langle b\right\rangle |\ge2$.
Partition $\mathcal{B}$ into cosets in the form $\left\langle b\right\rangle c=\{b^{n}\cdot c:\,n\in\mathbb{Z}\}$,
where $c\in\mathcal{B}$. Consider the entries of $\lambda_{\mathcal{B}}(b)(\mathbf{x})-\mathbf{x}$
with indices in $\left\langle b\right\rangle c$. The entry with index
$b^{n}\cdot c$ is $x_{b^{n-1}\cdot c}-x_{b^{n}\cdot c}$ for $n=0,\ldots,|\left\langle b\right\rangle |-1$
(note that $\lambda_{\mathcal{B}}(b)(\mathbf{x})$ is performing a
cyclic shift of entries within $\left\langle b\right\rangle c$).
It is straightforward to check that, if we only consider indices in
$\left\langle b\right\rangle c$, then the linear function $\{x_{b^{n}\cdot c}\}_{n=0,\ldots,|\left\langle b\right\rangle |-1}\mapsto\{(\lambda_{\mathcal{B}}(b)(\mathbf{x})-\mathbf{x})_{b^{n}\cdot c}\}_{n=0,\ldots,|\left\langle b\right\rangle |-1}$
has rank $|\left\langle b\right\rangle |-1$. In sum, $\mathbf{x}\mapsto\lambda_{\mathcal{B}}(b)(\mathbf{x})-\mathbf{x}$
has rank 
\[
\left(|\left\langle b\right\rangle |-1\right)\frac{|\mathcal{B}|}{|\left\langle b\right\rangle |}\ge\frac{|\mathcal{B}|}{2}
\]
since $|\left\langle b\right\rangle |\ge2$.
\end{IEEEproof}
\medskip{}

In this section, instead of showing a reduction from the uniform word
problem for finite monoids as in \cite{herrmann1995undecidability},
we will be using the uniform word problem for finite groups, which
is also undecidable \cite{slobodskoi1981undecidability}. We state
an equivalent form of the uniform word problem for finite groups:
Given $l,k\in\mathbb{Z}_{>0}$, $a_{i},b_{i},c_{i}\in\{1,\ldots,k\}$
for $i=1,\ldots,l$, determine whether the implication
\begin{equation}
\bigwedge_{i=1}^{l}\left(x_{a_{i}}\cdot x_{b_{i}}=x_{c_{i}}\right)\;\to\;x_{1}=e_{\mathcal{B}}\label{eq:impl_group}
\end{equation}
holds for all finite group $\mathcal{B}$ and all $k$-tuples $x_{1},\ldots,x_{k}\in\mathcal{B}$.
Note that the original word problem allows having inverse (e.g. $a^{-1}$)
appear in a word, though this can be emulated in \eqref{eq:impl_group}
by introducing an intermediate variable $\bar{a}$ satisfying $\bar{a}\cdot a=e$,
where $e$ is another variable satisfying $e\cdot e=e$ (which forces
$e$ to be the identity element), and hence we can use $\bar{a}$
in place of $a^{-1}$.

\medskip{}

\subsection{The Network for Enforcing Abelian Group}

We first prove a useful fact about random variables.
\begin{prop}
\label{prop:fcn_flip}If $X,Y,Z$ are finite random variables with
supports $\mathcal{X},\mathcal{Y},\mathcal{Z}$ respectively, satisfying
$X\perp Z$, $Z\stackrel{\iota}{\le}XY$, and $|\mathcal{Y}|\le|\mathcal{Z}|$,
then we have $|\mathcal{Y}|=|\mathcal{Z}|$ and $Y\stackrel{\iota}{\le}XZ$.
Moreover, if $Z$ is uniformly distributed, then $Y$ is uniformly
distributed and independent of $X$ as well.
\end{prop}
\begin{IEEEproof}
Let $g$ be a function such that $Z=g(X,Y)$. For any fixed $x$,
the set $\{g(x,y):\,y\in\mathcal{Y}\}$ has size at most $|\mathcal{Y}|$,
and it has size $|\mathcal{Y}|$ if and only if the function $y\mapsto g(x,y)$
is injective. Since $Z=g(x,Y)$ conditional on $X=x$, and the number
of possible values of $Z$ conditional on $X=x$ is $|\mathcal{Z}|\ge|\mathcal{Y}|$,
the number of possible values of $g(x,Y)$ must be $|\mathcal{Z}|$
as well, and hence $|\mathcal{Z}|=|\mathcal{Y}|$, and the function
$y\mapsto g(x,y)$ is injective, and hence it is bijective since $|\mathcal{Y}|=|\mathcal{Z}|$.
We can find function $h:\mathcal{X}\times\mathcal{Z}\to\mathcal{Y}$
such that $z\mapsto h(x,z)$ is the inverse of $y\mapsto g(x,y)$.
We have $Y=h(X,Z)$. If $Z$ is uniformly distributed, then $h(x,Z)$
is uniformly distributed as well for any fixed $x$ since $z\mapsto h(x,z)$
is bijective, implying that $Y=h(X,Z)$ is uniformly distributed and
independent of $X$.
\end{IEEEproof}
\medskip{}

Given a network and $\{M_{i}\}_{i=1,\ldots,n}$, $\{X_{f}\}_{f\in\tilde{\mathcal{L}}}$
for a subset $\tilde{\mathcal{L}}\subseteq\mathcal{L}$, we say that
$\{M_{i}\},\{X_{f}\}_{f\in\tilde{\mathcal{L}}}$ satisfy the coding
constraint if there exists $X_{f}$ for the remaining $f\in\mathcal{L}\backslash\tilde{\mathcal{L}}$
such that $\{M_{i}\},\{X_{f}\}_{f\in\mathcal{L}}$ satisfy the coding
constraint \eqref{eq:coding_cons}. We will use networks and subnetworks
to enforce various conditions on the messages and signals.

\begin{figure}
\begin{centering}
\includegraphics[scale=0.95]{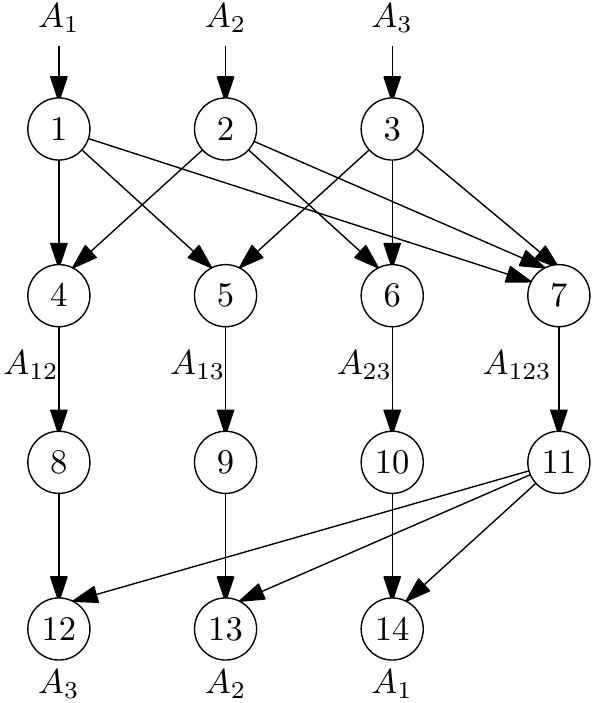}
\par\end{centering}
\caption{\label{fig:fnfnet}The base network, which is the network $\mathcal{N}_{1}$
in \cite[Fig. 1]{dougherty2006unachievability}.}
\end{figure}

We utilizes the network in \cite[Fig. 1]{dougherty2006unachievability}
as the base of our construction, which is given in Figure \ref{fig:fnfnet}
(also see \cite{dougherty2005insufficiency}). It was shown in \cite{dougherty2006unachievability}
that this network enforces the abelian group structure in \eqref{eq:abelian_label}.
 The following proposition follows from \cite[Prop. 5]{dougherty2006unachievability}
and Propositions \ref{prop:labeling} and \ref{prop:abelian_to_fnf},
though we include a short proof for the sake of completeness.

\begin{prop}
Given $A_{1},A_{2},A_{3}$ are independent and uniformly distributed
with cardinality $q$, and $\{A_{i}\}_{i\in\mathcal{E}}$ are random
variables with cardinality at most $q$. Then $\{A_{i}\}_{i\in\mathcal{E}}$
satisfy the Fano-non-Fano condition if and only if $\{A_{i}\}_{i\in\mathcal{E}}$
satisfy the coding constraint in the base network in Figure \ref{fig:fnfnet}
(where $A_{1},A_{2},A_{3}$ are source messages, and $A_{12},A_{13},A_{23},A_{123}$
are signals along edges).
\end{prop}
\begin{IEEEproof}
The ``only if'' direction is straightforward. For the ``if'' direction,
assume the coding constraint in the base network in Figure \ref{fig:fnfnet}
is satisfied. Since $A_{12}\perp A_{3}$ (by $A_{1}A_{2}\perp A_{3}$
and $A_{12}\stackrel{\iota}{\le}A_{1}A_{2}$) and $A_{3}\stackrel{\iota}{\le}A_{12}A_{123}$,
we have $A_{123}\stackrel{\iota}{\le}A_{12}A_{3}$ by Proposition
\ref{prop:fcn_flip}. Since $A_{1}A_{2}\perp A_{3}$ and $A_{3}\stackrel{\iota}{\le}A_{12}A_{123}\stackrel{\iota}{\le}A_{1}A_{2}A_{123}$,
we have $A_{1}A_{2}\perp A_{123}$ by Proposition \ref{prop:fcn_flip}.
Similarly, $A_{1}A_{3}\perp A_{123}$ and $A_{2}A_{3}\perp A_{123}$.
Since $A_{123}\perp A_{3}$ and $A_{3}\stackrel{\iota}{\le}A_{123}A_{12}$,
we have $A_{12}\stackrel{\iota}{\le}A_{123}A_{3}$ and $A_{123}\perp A_{12}$
by Proposition \ref{prop:fcn_flip}. Hence $\mathrm{tri}(A_{12},A_{3},A_{123})$
holds. Since $A_{1}A_{3}\perp A_{2}$ and $A_{2}\stackrel{\iota}{\le}A_{13}A_{123}\stackrel{\iota}{\le}A_{1}A_{3}A_{12}$,
we have $A_{1}A_{3}\perp A_{12}$ by Proposition \ref{prop:fcn_flip}.
Since $A_{1}\perp A_{12}$ and $A_{12}\stackrel{\iota}{\le}A_{1}A_{2}$,
we have $A_{2}\stackrel{\iota}{\le}A_{1}A_{12}$ by Proposition \ref{prop:fcn_flip}.
Similarly, $A_{1}\stackrel{\iota}{\le}A_{2}A_{12}$. Hence $\mathrm{tri}(A_{1},A_{2},A_{12})$
holds. The result follows from Proposition \ref{prop:fnf_equiv}.
\end{IEEEproof}
\medskip{}

\begin{figure}
\begin{centering}
\includegraphics[scale=0.95]{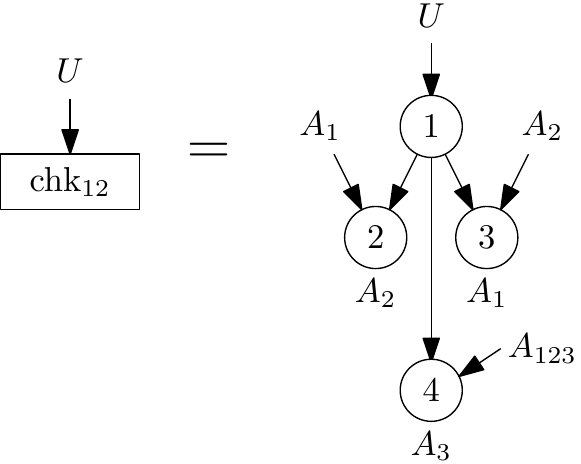}
\par\end{centering}
\caption{\label{fig:chk12}The $\mathrm{chk}_{12}$ subnetwork, which checks
whether $U\stackrel{\iota}{=}A_{12}$. }
\end{figure}

If we want to enforce that the signal along an edge is $A_{1}$, we
can simply require the ending node to decode $A_{1}$. However, if
we want to enforce that the signal is $A_{12}$, then we require a
subnetwork given in Figure \ref{fig:chk12}, called the $\mathrm{chk}_{12}$
subnetwork. The subnetwork has two inputs $U$ and $A_{123}$ (which
is always chosen to be $A_{123}$ from the base network in Figure
\ref{fig:fnfnet}, and is omitted later in network diagrams), and
checks whether $U\stackrel{\iota}{=}A_{12}$. More explicitly, the
subnetwork checks the condition
\begin{align*}
 & \mathrm{chk}_{12}(\{A_{i}\}_{i},U):\\
 & A_{1}\stackrel{\iota}{\le}A_{2}U\,\wedge\,A_{2}\stackrel{\iota}{\le}A_{1}U\,\wedge\,A_{3}\stackrel{\iota}{\le}A_{123}U.
\end{align*}

\medskip{}

We now show that the $\mathrm{chk}_{12}$ subnetwork can check if
a signal has the same information as $A_{12}$
\begin{prop}
Given $\{A_{i}\}_{i\in\mathcal{E}}$ satisfying the Fano-non-Fano
condition \eqref{eq:fnf}, each with cardinality $q$, and $U$ is
a random variable with cardinality at most $q$. We have $\mathrm{chk}_{12}(\{A_{i}\}_{i},U)$
if and only if $U\stackrel{\iota}{=}A_{12}$.
\end{prop}
\begin{IEEEproof}
The ``if'' direction is straightforward. For the ``only if'' direction,
assume $A_{1}\stackrel{\iota}{\le}A_{2}U\,\wedge\,A_{2}\stackrel{\iota}{\le}A_{1}U\,\wedge\,A_{3}\stackrel{\iota}{\le}A_{123}U$.
Since $A_{2}\perp A_{1}$ and $A_{1}\stackrel{\iota}{\le}A_{2}U$,
we have $U\stackrel{\iota}{\le}A_{2}A_{1}$ and $U\perp A_{2}$ by
Proposition \ref{prop:fcn_flip}. Hence $\mathrm{tri}(A_{1},A_{2},U)$
holds. Since $A_{2}A_{3}\perp A_{1}$ and $A_{1}\stackrel{\iota}{\le}A_{2}A_{3}U$,
we have $A_{2}A_{3}\perp U$ by Proposition \ref{prop:fcn_flip}.
Since $U\perp A_{3}$ and $A_{3}\stackrel{\iota}{\le}UA_{123}$, we
have $A_{123}\stackrel{\iota}{\le}UA_{3}$ by Proposition \ref{prop:fcn_flip}.
Since $A_{123}\perp A_{3}$ and $A_{3}\stackrel{\iota}{\le}A_{123}U$,
we have $U\stackrel{\iota}{\le}A_{123}A_{3}$ and $A_{123}\perp U$
by Proposition \ref{prop:fcn_flip}. Hence $\mathrm{tri}(U,A_{3},A_{123})$
holds. Therefore, we have $\mathrm{fnf}(A_{1},A_{2},A_{3},U,A_{13},A_{23},A_{123})$
by Proposition \ref{prop:fnf_equiv}.

By Proposition \ref{prop:labeling}, let $(\mathcal{A},\{\theta_{i}\}_{i\in\mathcal{E}})$
be an abelian group labeling of $\{A_{i}\}$, and $(\tilde{\mathcal{A}},\{\tilde{\theta}_{i}\}_{i\in\mathcal{E}})$
be an abelian group labeling of $(A_{1},A_{2},A_{3},U,A_{13},A_{23},A_{123})$.
We use $+$ for the group operation of $\mathcal{A}$, and $\tilde{+}$
for the group operation of $\tilde{\mathcal{A}}$. We have $\theta_{1}A_{1}+\theta_{2}A_{2}+\theta_{3}A_{3}=\theta_{123}A_{123}$
(write $\theta_{1}A_{1}=\theta_{1}(A_{1})$ for brevity). Conditional
on the event $\tilde{\theta}_{3}A_{3}=0$, we have
\[
A_{123}=\theta_{123}^{-1}\left(\theta_{1}A_{1}+\theta_{2}A_{2}+\theta_{3}\tilde{\theta}_{3}^{-1}0\right).
\]
Combining this with $\tilde{\theta}_{1}A_{1}\tilde{+}\tilde{\theta}_{2}A_{2}=\tilde{\theta}_{123}A_{123}$
conditional on $\tilde{\theta}_{3}A_{3}=0$, and noting that $(A_{1},A_{2})$
can be any pair of values conditional on $\tilde{\theta}_{3}A_{3}=0$,
we have, for any $a_{1},a_{2}$,
\[
\tilde{\theta}_{1}a_{1}\tilde{+}\tilde{\theta}_{2}a_{2}=\tilde{\theta}_{123}\theta_{123}^{-1}\left(\theta_{1}a_{1}+\theta_{2}a_{2}+\theta_{3}\tilde{\theta}_{3}^{-1}0\right).
\]
Hence $\tilde{\theta}_{12}U=\tilde{\theta}_{1}A_{1}\tilde{+}\tilde{\theta}_{2}A_{2}$
contains the same information as $\theta_{12}A_{12}=\theta_{1}A_{1}+\theta_{2}A_{2}$.
\end{IEEEproof}
\medskip{}

\subsection{Enforcing Automorphisms via Subnetworks}

\begin{figure}
\begin{centering}
\includegraphics[scale=0.95]{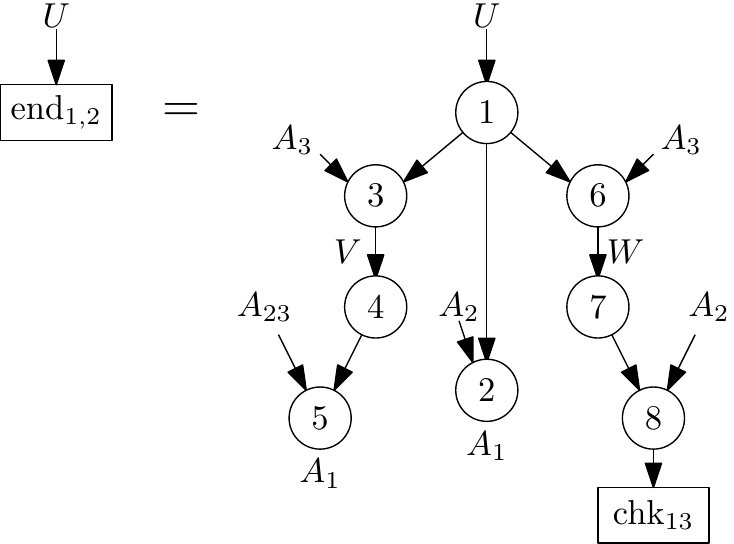}
\par\end{centering}
\caption{\label{fig:endnet}The $\mathrm{end}_{12}$ subnetwork, which checks
whether $U$ corresponds to an endomorphism. Note that it uses the
$\mathrm{chk}_{13}$ subnetwork defined in Figure \ref{fig:fnfnet}.
While Figure \ref{fig:fnfnet} defines the $\mathrm{chk}_{12}$ subnetwork,
it can be converted to a $\mathrm{chk}_{13}$ subnetwork by swapping
the indices $2$ and $3$.}
\end{figure}

First we introduce a subnetwork that checks the condition $\mathrm{end}_{1,2}(\{A_{i}\},U)$
in \eqref{eq:end}. The subnetwork is given in Figure \ref{fig:endnet}.

\begin{prop}
Given $\{A_{i}\}_{i\in\mathcal{E}}$ satisfying the Fano-non-Fano
condition \eqref{eq:fnf}, each with cardinality $q$, and $U$ is
a random variable with cardinality at most $q$. We have $\mathrm{end}_{1,2}(\{A_{i}\},U)$
if and only if $\{A_{i}\},U$ satisfy the coding constraint in the
$\mathrm{end}_{12}$ subnetwork in Figure \ref{fig:endnet}.\footnote{Technically we have only defined the notion of coding constraint of
a network, but not a subnetwork. The only difference between a network
and a subnetwork is that a subnetwork can have inputs that are not
source messages, and those inputs may not be independent. The exact
same notion of coding constraint \eqref{eq:coding_cons} can be applied
to subnetworks.}
\end{prop}
\begin{IEEEproof}
The ``only if'' direction follows directly from the definition.
We now prove the ``if'' direction. Since $A_{2}\perp A_{1}$ and
$A_{1}\stackrel{\iota}{\le}A_{2}U$, we have $U\stackrel{\iota}{\le}A_{2}A_{1}$
(and hence $U\stackrel{\iota}{=}A_{1}|A_{2}$), $U$ is uniform with
cardinality $q$, and $U\perp A_{2}$ by Proposition \ref{prop:fcn_flip}.
Similarly, $V\stackrel{\iota}{=}A_{1}|A_{23}$, $W\stackrel{\iota}{=}A_{13}|A_{2}$,
and $V,W$ are uniform with cardinality $q$. Since $A_{2}A_{3}\perp A_{1}$
and $A_{1}\stackrel{\iota}{\le}A_{23}V\stackrel{\iota}{\le}A_{2}A_{3}V$,
we have $A_{2}A_{3}\perp V$ by Proposition \ref{prop:fcn_flip}.
Since $A_{3}\perp V$ and $V\stackrel{\iota}{\le}A_{3}U$, we have
$U\stackrel{\iota}{\le}A_{3}V$ (and hence $U\stackrel{\iota}{=}V|A_{3}$).
Since $A_{2}A_{3}\perp A_{13}$ and $A_{13}\stackrel{\iota}{\le}A_{2}W\stackrel{\iota}{\le}A_{2}A_{3}W$,
we have $A_{2}A_{3}\perp W$ by Proposition \ref{prop:fcn_flip}.
Since $A_{3}\perp W$ and $W\stackrel{\iota}{\le}A_{3}U$, we have
$U\stackrel{\iota}{\le}A_{3}W$ (and hence $U\stackrel{\iota}{=}W|A_{3}$).
\end{IEEEproof}
\medskip{}

\begin{figure}
\begin{centering}
\includegraphics[scale=0.95]{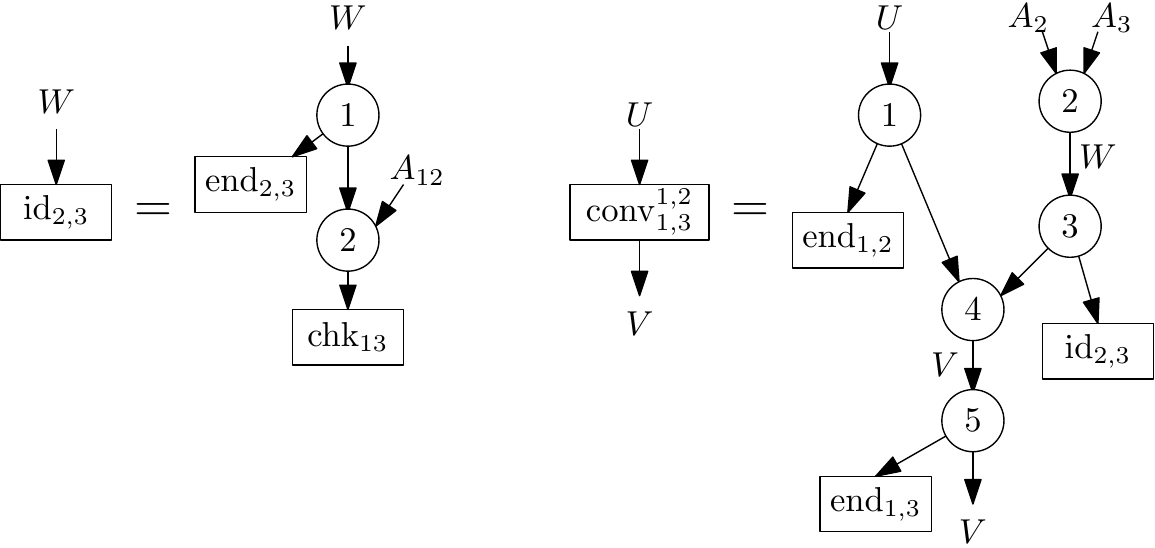}
\par\end{centering}
\caption{\label{fig:convnet}Left: The $\mathrm{id}_{2,3}$ subnetwork, which
checks whether $W$ satisfies $\mathrm{end}_{2,3}(\{A_{i}\},W)$ and
corresponds to the identity endomorphism (i.e., $W\stackrel{\iota}{=}A_{2}-A_{3}$).
Right: The $\mathrm{conv}_{1,3}^{1,2}$ subnetwork, which converts
$U$ satisfying $\mathrm{end}_{1,2}(\{A_{i}\},U)$ to $V$ satisfying
$\mathrm{end}_{1,3}(\{A_{i}\},V)$.}
\end{figure}

Next, we introduce a subnetwork for the condition $\mathrm{conv}_{1,3}^{1,2}(\{A_{i}\},U,V)$
in \eqref{eq:conv1213}. The subnetwork is given in Figure \ref{fig:convnet}.
Note that $U$ is an input to the subnetwork, and $V$ is an output,
and $U,V$ satisfy the coding constraint if and only if $\mathrm{conv}_{1,3}^{1,2}(\{A_{i}\},U,V)$
hold. We also introduce the $\mathrm{id}_{2,3}$ subnetwork that checks
whether $W$ satisfies $\mathrm{end}_{2,3}(\{A_{i}\},W)$ and corresponds
to the identity endomorphism (i.e., $W\stackrel{\iota}{=}A_{2}-A_{3}$).
Figure \ref{fig:convnet} can be obtained directly from \eqref{eq:conv1213},
so we omit the proof. The $\mathrm{conv}_{3,2}^{1,2}$ subnetwork
for \eqref{eq:conv1232} can be obtained similarly.

\begin{figure}
\begin{centering}
\includegraphics[scale=0.95]{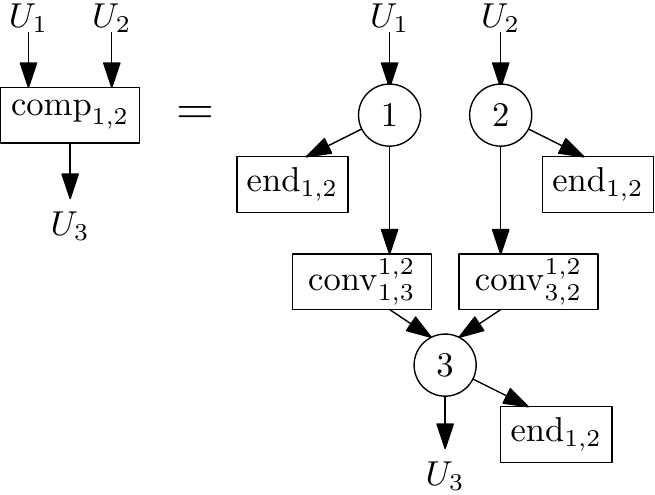}
\par\end{centering}
\caption{\label{fig:compnet}The $\mathrm{comp}_{1,2}$ subnetwork, where the
inputs are $U_{1},U_{2}$ corresponding to endomorphisms $g_{1},g_{2}$,
and outputs $U_{3}$ corresponding to the endomorphism $g_{3}=g_{1}\cdot g_{2}$.}
\end{figure}

We then introduce a subnetwork for the condition $\mathrm{comp}_{1,2}(\{A_{i}\},U_{1},U_{2},U_{3})$
in \eqref{prop:comp_12}. The subnetwork is given in Figure \ref{fig:compnet}.
Figure \ref{fig:compnet} can be obtained directly from \eqref{prop:comp_12},
so we omit the proof. 

\begin{figure}
\begin{centering}
\includegraphics[scale=0.95]{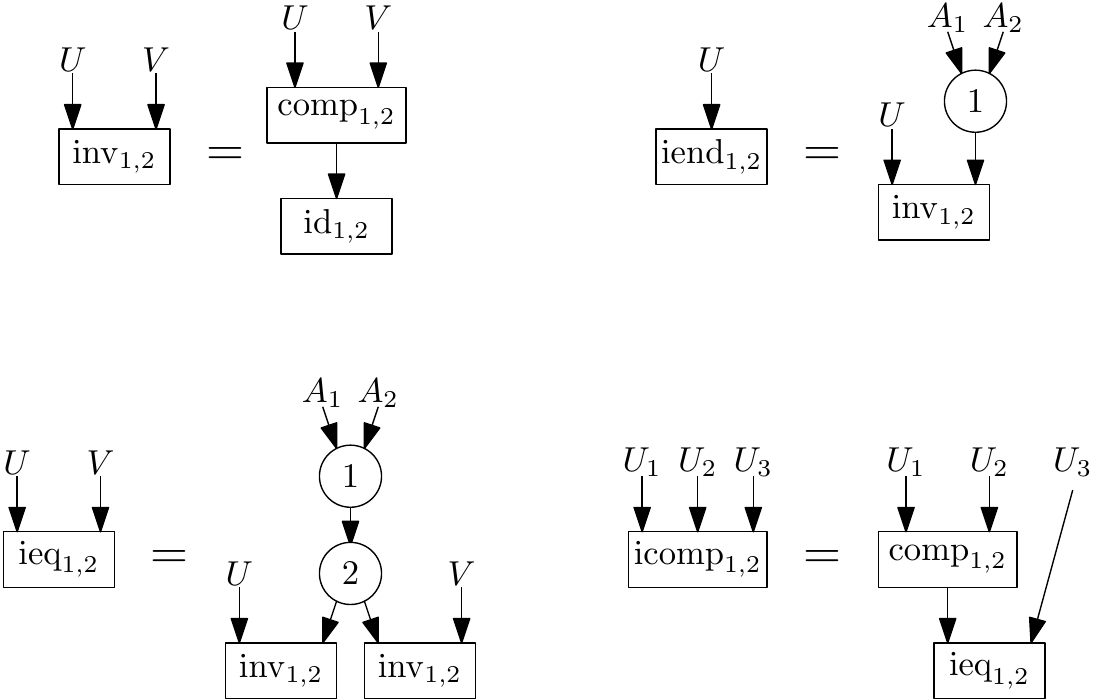}
\par\end{centering}
\caption{\label{fig:invnet}Top left: The $\mathrm{inv}_{1,2}$ subnetwork,
which checks whether the endomorphisms corresponding to $U,V$ are
automorphisms and are inverses of each other. Top right: The $\mathrm{iend}_{1,2}$
subnetwork, which checks whether the endomorphisms corresponding to
$U$ is an automorphism. Bottom left: The $\mathrm{ieq}_{1,2}$ subnetwork,
which checks whether the endomorphisms corresponding to $U,V$ are
automorphisms and are the same. Bottom right: The $\mathrm{icomp}_{1,2}$
subnetwork, which checks whether the endomorphisms $g_{1},g_{2},g_{3}$
corresponding to $U_{1},U_{2},U_{3}$ are automorphisms and satisfy
$g_{3}=g_{1}\cdot g_{2}$.}
\end{figure}

The problem of the subnetwork in Figure \ref{fig:compnet} is that
$U_{3}$ is an output. This is undesirable since for each random variable
$U_{3}$, we will only be able to enforce one instance of $\mathrm{comp}_{1,2}(\{A_{i}\},U_{1},U_{2},U_{3})$,
since $U_{3}$ can only appear once as the output of a subnetwork.
To be able to enforce many instances of $\mathrm{comp}_{1,2}$, we
will require the four subnetworks given in Figure \ref{fig:invnet}:
1) the $\mathrm{inv}_{1,2}$ subnetwork, which checks whether the
endomorphisms corresponding to $U,V$ are automorphisms and are inverses
of each other (by checking whether their composition is the identity
endomorphism, note that an endomorphism on a finite group having a
left or right inverse is sufficient for it to be an automorphism);
2) the $\mathrm{iend}_{1,2}$ subnetwork, which checks whether the
endomorphisms corresponding to $U$ is an automorphism (by checking
whether an inverse exists); 3) the $\mathrm{ieq}_{1,2}$ subnetwork,
which checks whether the endomorphisms corresponding to $U,V$ are
automorphisms and are the same (by checking whether they have the
same inverse); and 4) the $\mathrm{icomp}_{1,2}$ subnetwork, which
checks whether the endomorphisms $g_{1},g_{2},g_{3}$ corresponding
to $U_{1},U_{2},U_{3}$ are automorphisms and satisfy $g_{3}=g_{1}\cdot g_{2}$
(note that $g_{1}\cdot g_{2}$ being an automorphism implies that
$g_{1}$ and $g_{2}$ are automorphisms). 

The main reason we use the uniform word problem for finite groups
instead of finite monoids is that we require the inverse in order
to check for equality of automorphisms, and hence to have a subnetwork
that checks for composition which takes $U_{1},U_{2},U_{3}$ as inputs.
Designing a subnetwork that checks for equality on general endomorphisms
(not necessarily automorphisms) does not appear to be straightforward.

\begin{figure}
\begin{centering}
\includegraphics[scale=0.95]{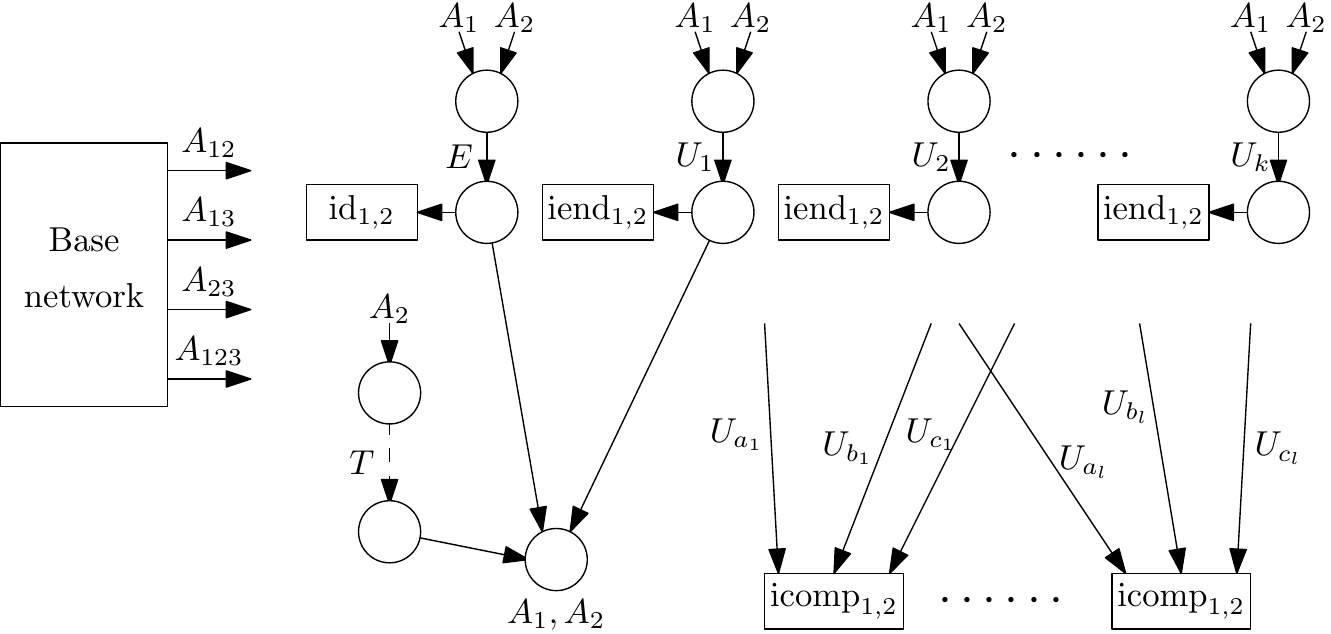}
\par\end{centering}
\caption{\label{fig:finalnet}The final network, which contains the the base
network in Figure \ref{fig:fnfnet}. Each $U_{j}$ ($j=1,\ldots,k$)
is input to an $\mathrm{iend}_{1,2}$ subnetwork. There are $l$ $\mathrm{icomp}_{1,2}$
subnetwork, where the $j$-th one is connected to $U_{a_{j}},U_{b_{j}},U_{c_{j}}$.
All solid edges, as well as all edges in all subnetworks and the the
base network, are assumed to be two parallel edges. Only the dashed
edge carrying $T$ is a single edge.}
\end{figure}

The final network is given in Figure \ref{fig:finalnet}, which contains
the the base network in Figure \ref{fig:fnfnet} (it has edges sending
$A_{12},A_{13},A_{23},A_{123}$ to the other subnetworks, which are
omitted for clarity). Each $U_{j}$ ($j=1,\ldots,k$) is input to
an $\mathrm{iend}_{1,2}$ subnetwork. There are $l$ $\mathrm{icomp}_{1,2}$
subnetwork, where the $j$-th one is connected to $U_{a_{j}},U_{b_{j}},U_{c_{j}}$.
All solid edges, as well as all edges in all subnetworks and the base
 network, are assumed to be two parallel edges (together they can
transmit $q^{2}$ possible values, and we can regard the alphabet
size of these components to be $q^{2}$). The source messages are
also duplicated accordingly, i.e., we let $A_{i}=A'_{i}A''_{i}$ for
$i=1,2,3$, where each of $A_{i}',A''_{i}$ has cardinality $q$,
so we can treat $A_{i}$ as a source message with alphabet size $q^{2}$.
The source messages are $A_{1},A_{2},A_{3}$, or more precisely, $A'_{1},A''_{1},A'_{2},A''_{2},A'_{3},A''_{3}$.
Only the dashed edge carrying $T$ (which is a function of $A_{2}$)
is a single edge that can only transmit $q$ different values. 

The network is designed to be solvable if and only if there exists
$q\in\mathbb{Z}_{\ge2}$, and random variables $A_{i}$ ($i\in\mathcal{E}$)
with cardinality at most $q^{2}$, $U_{i}$ ($i=1,\ldots,k$) with
cardinality at most $q^{2}$, $E$ with cardinality at most $q^{2}$,
and $T$ with cardinality at most $q$, satisfying that $A_{1},A_{2},A_{3}$
are independent and uniformly distributed with cardinality $q^{2}$,
and

\begin{align}
 & \bigwedge_{j=1}^{k}\mathrm{iend}_{1,2}(\{A_{i}\},U_{j})\nonumber \\
 & \wedge\,\bigwedge_{j=1}^{l}\mathrm{icomp}_{1,2}(\{A_{i}\},U_{a_{j}},U_{b_{j}},U_{c_{j}})\nonumber \\
 & \wedge\,\mathrm{id}_{1,2}(\{A_{i}\},E)\,\wedge\,T\stackrel{\iota}{\le}A_{2}\,\wedge\,A_{1}A_{2}\stackrel{\iota}{\le}U_{1}ET.\label{eq:impl_network}
\end{align}

We complete the proof of Theorems \ref{thm:netcode} and \ref{thm:netcode_vec}
by showing that the network is not solvable if and only if \eqref{eq:impl_group}
holds.
\begin{prop}
For any fixed $c\in\mathbb{Z}_{\ge2}$, the following are equivalent:
\begin{itemize}
\item The implication \eqref{eq:impl_group} holds for all finite group
$\mathcal{B}$ and all $x_{1},\ldots,x_{k}\in\mathcal{B}$. 
\item The network in Figure \ref{fig:finalnet} is unsolvable. 
\item The network in Figure \ref{fig:finalnet} is unsolvable with alphabet
size $q=c^{m}$ for all $m\in\mathbb{Z}_{\ge1}$. 
\end{itemize}
\end{prop}
\begin{IEEEproof}
First, we show that if the implication \eqref{eq:impl_group} does
not hold, then the network is solvable, and also solvable for some
$q=c^{m}$. Assume the implication \eqref{eq:impl_group} does not
hold. There exists a finite group $\mathcal{B}$ and $x_{1},\ldots,x_{k}\in\mathcal{B}$
satisfying the left hand side of \eqref{eq:impl_group}, but not the
right hand side, i.e., we have $x_{1}\neq e_{\mathcal{B}}$. Fix any
prime $p$ and finite field $\mathbb{F}$ with $|\mathbb{F}|=p^{2}$.
Consider the left regular representation $\lambda_{\mathcal{B}}:\mathcal{B}\to\mathrm{GL}(\mathbb{F}^{\mathcal{B}})$.
By Proposition \ref{prop:abelian_to_fnf}, let $\{A_{i}\}$ satisfy
$\mathrm{fnf}(\{A_{i}\})$ such that $(\mathbb{F}^{\mathcal{B}},\{\theta_{i}\}_{i\in\mathcal{E}})$
is an abelian group labeling, where $\mathbb{F}^{\mathcal{B}}$ is
treated as an abelian group under addition here. Assume $\theta_{i}$
are identity functions, so $A_{i}\in\mathbb{F}^{\mathcal{B}}$. Note
that the cardinality of $A_{i}$ is $p^{2|\mathcal{B}|}\ge4$. Let
$U_{j}$ satisfy $\mathrm{iend}_{1,2}(\{A_{i}\},U_{j})$ corresponding
to the endomorphism $\lambda_{\mathcal{B}}(x_{j})$ (by Proposition
\ref{prop:end_to_fnf} and that $\lambda_{\mathcal{B}}(x_{j})$ is
an automorphism with an inverse). We have $\mathrm{iend}_{1,2}(\{A_{i}\},U_{j})$
and $\mathrm{icomp}_{1,2}(\{A_{i}\},U_{a_{j}},U_{b_{j}},U_{c_{j}})$
by the left hand side of \eqref{eq:impl_group}. Since $x_{1}\neq e_{\mathcal{B}}$,
by Proposition \ref{prop:non_id}, the linear function $\lambda_{\mathcal{B}}(x_{1})-\mathrm{id}_{\mathbb{F}^{\mathcal{B}}}$
has rank at least $\lceil|\mathcal{B}|/2\rceil$, and hence there
exists a linear function $t:\mathbb{F}^{\mathcal{B}}\to\mathbb{F}^{\lfloor|\mathcal{B}|/2\rfloor}$
such that the function $\mathbf{x}\mapsto(\lambda_{\mathcal{B}}(x_{1})(\mathbf{x})-\mathbf{x},t(\mathbf{x}))$
is injective. Let $T=t(A_{2})$ with cardinality at most $|\mathbb{F}^{\lfloor|\mathcal{B}|/2\rfloor}|\le p^{|\mathcal{B}|}$.
Then $A_{2}$ can be deduced from $U_{1}=A_{1}-\lambda_{\mathcal{B}}(x_{1})(A_{2})$,
$E=A_{1}-A_{2}$ (corresponds to the identity endomorphism) and $T=t(A_{2})$
by considering $(E-U_{1},T)=(\lambda_{\mathcal{B}}(x_{1})(A_{2})-A_{2},t(A_{2}))$,
and $A_{1}=E+A_{2}$ can be deduced as well. Hence \eqref{eq:impl_network}
is satisfied with $q=p^{|\mathcal{B}|}$, and the network is solvable.

To show that the network is solvable for some $q=c^{m}$, let $c=\prod_{i=1}^{n}p_{i}$,
where $p_{i}$ are primes (possibly with duplicates). We have shown
that the network is solvable with $q=p_{i}^{|\mathcal{B}|}$. By combining
these codes together, the network is solvable with $q=\prod_{i=1}^{n}p_{i}^{|\mathcal{B}|}=c^{|\mathcal{B}|}$.

Next, we show that if the implication \eqref{eq:impl_group} holds,
then the network is not solvable, and hence not solvable for any $q=c^{m}$.
Assume the implication \eqref{eq:impl_group} holds. Assume the contrary
that the network is solvable, and \eqref{eq:impl_network} holds.
Fix any abelian group labeling $(\mathcal{A},\{\theta_{i}\}_{i\in\mathcal{E}})$.
By $\mathrm{iend}_{1,2}(\{A_{i}\},U_{j})$, we can find the endomorphism
$x_{j}\in\mathrm{End}(\mathcal{A})$ corresponding to $U_{j}$ (Proposition
\ref{prop:end_label}), which is actually an automorphism since $\mathrm{iend}_{1,2}(\{A_{i}\},U_{j})$
holds. By $\mathrm{icomp}_{1,2}(\{A_{i}\},U_{a_{j}},U_{b_{j}},U_{c_{j}})$,
we have $x_{a_{j}}\cdot x_{b_{j}}=x_{c_{j}}$. Applying \eqref{eq:impl_group}
on the automorphism group $\mathrm{Aut}(\mathcal{A})$, we have $x_{1}=e_{\mathrm{Aut}(\mathcal{A})}$.
Hence $U_{1}$ contains the same information as $A_{1}-A_{2}$, which
is also the same information as $E$. The tuple $(U_{1},E,T)$ can
have at most $q^{3}$ different values, and cannot be used to deduce
$(A_{1},A_{2})$ which has $q^{4}$ different values, giving a contradiction.
Hence the network is not solvable.
\end{IEEEproof}
\medskip{}

As a corollary of Theorem \ref{thm:netcode} and the observation in
\cite{lehman2005network} that network coding would be decidable if
there is a computable upper bound on the alphabet size, the minimum
alphabet size needed to solve a network is not upper-bounded by any
computable function.\footnote{A computable function is a function that can be computed by an algorithm
(or a Turing machine).} We include a precise statement and a proof for the sake of completeness.
For a solvable network, its minimum alphabet size is the smallest
$q$ such that the network is solvable with alphabet size $q$. Let
$q_{\max}(m)$ be the maximum of the minimum alphabet sizes of solvable
networks with at most $m$ nodes, $m$ edges and $m$ source messages.
Note that $q_{\max}(m)$ is finite since there are finitely many networks
with at most $m$ nodes/edges/messages. Then $q_{\max}(m)$ is not
upper-bounded by any computable function.
\begin{cor}
\label{cor:computable}There does not exist any computable function
$f:\mathbb{Z}_{\ge1}\to\mathbb{Z}_{\ge2}$ satisfying that $q_{\max}(m)=O(f(m))$
as $m\to\infty$.
\end{cor}
\begin{IEEEproof}
Assume the contrary that there is a computable function $f$ such
that $q_{\max}(m)=O(f(m))$. Let $q_{\max}(m)\le cf(m)$ for $m\ge m_{0}$.
Let $g(m):=\lceil c\rceil f(m)$ for $m\ge m_{0}$, and $g(m)=q_{\max}(m)$
for $m<m_{0}$. Note that $q_{\max}(m)\le g(m)$, and $g$ is computable
since $m\mapsto\lceil c\rceil f(m)$ is computable, and a computable
function is still computable after changing finitely many values.
Consider an algorithm that, given a network, compute $m$ as the maximum
among the number of nodes, number of edges and number of source messages,
exhaust all coding schemes (encoding functions at each node) with
alphabet size at most $g(m)$, and output ``solvable'' if any coding
scheme satisfies the decoding constraint, and output ``unsolvable''
otherwise.\footnote{This is the strategy used in \cite{lehman2005network}.}
If the network is solvable, since $q_{\max}(m)\le g(m)$, the algorithm
finds a working coding scheme, and correctly outputs ``solvable''.
If the network is unsolvable, then the algorithm cannot find any working
coding scheme, and correctly outputs ``unsolvable''. We have found
an algorithm that determines whether a network is solvable, contradicting
Theorem \ref{thm:netcode}.
\end{IEEEproof}

\section{Acknowledgement}

This work was supported in part by the Hong Kong Research Grant Council
Grant ECS No. CUHK 24205621, and the Direct Grant for Research, The
Chinese University of Hong Kong (Project ID: 4055133).  The author
would like to thank an anonymous reviewer of another paper by the
author, who raised the question whether the techniques in \cite{herrmann1995undecidability}
can be applied to prove undecidability results on probabilistic conditional
independence.

\medskip{}

\[
\]

\bibliographystyle{IEEEtran}
\bibliography{ref}

\begin{thebibliography}{10}
\providecommand{\url}[1]{#1}
\csname url@samestyle\endcsname
\providecommand{\newblock}{\relax}
\providecommand{\bibinfo}[2]{#2}
\providecommand{\BIBentrySTDinterwordspacing}{\spaceskip=0pt\relax}
\providecommand{\BIBentryALTinterwordstretchfactor}{4}
\providecommand{\BIBentryALTinterwordspacing}{\spaceskip=\fontdimen2\font plus
\BIBentryALTinterwordstretchfactor\fontdimen3\font minus
  \fontdimen4\font\relax}
\providecommand{\BIBforeignlanguage}[2]{{%
\expandafter\ifx\csname l@#1\endcsname\relax
\typeout{** WARNING: IEEEtran.bst: No hyphenation pattern has been}%
\typeout{** loaded for the language `#1'. Using the pattern for}%
\typeout{** the default language instead.}%
\else
\language=\csname l@#1\endcsname
\fi
#2}}
\providecommand{\BIBdecl}{\relax}
\BIBdecl

\bibitem{ahlswede2000network}
R.~Ahlswede, N.~Cai, S.-Y. Li, and R.~W. Yeung, ``Network information flow,''
  \emph{IEEE Transactions on information theory}, vol.~46, no.~4, pp.
  1204--1216, 2000.

\bibitem{li2003linear}
S.-Y. Li, R.~W. Yeung, and N.~Cai, ``Linear network coding,'' \emph{IEEE
  transactions on information theory}, vol.~49, no.~2, pp. 371--381, 2003.

\bibitem{jaggi2005polynomial}
S.~Jaggi, P.~Sanders, P.~A. Chou, M.~Effros, S.~Egner, K.~Jain, and L.~M.
  Tolhuizen, ``Polynomial time algorithms for multicast network code
  construction,'' \emph{IEEE Transactions on Information Theory}, vol.~51,
  no.~6, pp. 1973--1982, 2005.

\bibitem{harvey2005deterministic}
N.~J.~A. Harvey, ``Deterministic network coding by matrix completion,'' Ph.D.
  dissertation, Massachusetts Institute of Technology, 2005.

\bibitem{li2005achieving}
Z.~Li, B.~Li, D.~Jiang, and L.~C. Lau, ``On achieving optimal throughput with
  network coding,'' in \emph{Proceedings IEEE 24th Annual Joint Conference of
  the IEEE Computer and Communications Societies.}, vol.~3.\hskip 1em plus
  0.5em minus 0.4em\relax IEEE, 2005, pp. 2184--2194.

\bibitem{ho2006random}
T.~Ho, M.~M{\'e}dard, R.~Koetter, D.~R. Karger, M.~Effros, J.~Shi, and
  B.~Leong, ``A random linear network coding approach to multicast,''
  \emph{IEEE Transactions on Information Theory}, vol.~52, no.~10, pp.
  4413--4430, 2006.

\bibitem{lehman2005network}
A.~R. Lehman, ``Network coding,'' Ph.D. dissertation, Massachusetts Institute
  of Technology, 2005.

\bibitem{langberg2006encoding}
M.~Langberg, A.~Sprintson, and J.~Bruck, ``The encoding complexity of network
  coding,'' \emph{IEEE Transactions on Information Theory}, vol.~52, no.~6, pp.
  2386--2397, 2006.

\bibitem{yao2009network}
H.~Yao and E.~Verbin, ``Network coding is highly non-approximable,'' in
  \emph{2009 47th Annual Allerton Conference on Communication, Control, and
  Computing (Allerton)}.\hskip 1em plus 0.5em minus 0.4em\relax IEEE, 2009, pp.
  209--213.

\bibitem{langberg2011hardness}
M.~Langberg and A.~Sprintson, ``On the hardness of approximating the network
  coding capacity,'' \emph{IEEE Transactions on Information Theory}, vol.~57,
  no.~2, pp. 1008--1014, 2011.

\bibitem{dougherty2011network}
R.~Dougherty, C.~Freiling, and K.~Zeger, ``Network coding and matroid theory,''
  \emph{Proceedings of the IEEE}, vol.~99, no.~3, pp. 388--405, 2011.

\bibitem{kuhne2019representability}
L.~K{\"u}hne and G.~Yashfe, ``Representability of matroids by c-arrangements is
  undecidable,'' \emph{arXiv preprint arXiv:1912.06123}, 2019.

\bibitem{dougherty2009undecidable}
R.~Dougherty, ``Is network coding undecidable?'' in \emph{Applications of
  Matroid Theory and Combinatorial Optimization to Information and Coding
  Theory}, 2009.

\bibitem{albert1992undecidability}
D.~Albert, R.~Baldinger, and J.~Rhodes, ``Undecidability of the identity
  problem for finite semigroups,'' \emph{The Journal of symbolic logic},
  vol.~57, no.~1, pp. 179--192, 1992.

\bibitem{li2021netcode}
C.~T. Li, ``The undecidability of network coding with some fixed-size messages
  and edges,'' \emph{arXiv preprint arXiv:2109.08991}, 2021.

\bibitem{cannons2006network}
J.~Cannons, R.~Dougherty, C.~Freiling, and K.~Zeger, ``Network routing
  capacity,'' \emph{IEEE Transactions on Information Theory}, vol.~52, no.~3,
  pp. 777--788, 2006.

\bibitem{dougherty2008linear}
R.~Dougherty, C.~Freiling, and K.~Zeger, ``Linear network codes and systems of
  polynomial equations,'' \emph{IEEE Transactions on Information Theory},
  vol.~54, no.~5, pp. 2303--2316, 2008.

\bibitem{langberg2009multiple}
M.~Langberg and M.~M{\'e}dard, ``On the multiple unicast network coding,
  conjecture,'' in \emph{2009 47th Annual Allerton Conference on Communication,
  Control, and Computing (Allerton)}.\hskip 1em plus 0.5em minus 0.4em\relax
  IEEE, 2009, pp. 222--227.

\bibitem{bassoli2013network}
R.~Bassoli, H.~Marques, J.~Rodriguez, K.~W. Shum, and R.~Tafazolli, ``Network
  coding theory: A survey,'' \emph{IEEE Communications Surveys \& Tutorials},
  vol.~15, no.~4, pp. 1950--1978, 2013.

\bibitem{huang2013secure}
W.~Huang, T.~Ho, M.~Langberg, and J.~Kliewer, ``On secure network coding with
  uniform wiretap sets,'' in \emph{2013 International Symposium on Network
  Coding (NetCod)}.\hskip 1em plus 0.5em minus 0.4em\relax IEEE, 2013, pp.
  1--6.

\bibitem{gomez2014network}
A.~G{\'o}mez, C.~Mej{\'\i}a, and J.~A. Montoya, ``Network coding and the model
  theory of linear information inequalities,'' in \emph{2014 International
  Symposium on Network Coding (NetCod)}.\hskip 1em plus 0.5em minus 0.4em\relax
  IEEE, 2014, pp. 1--6.

\bibitem{dawid1979conditional}
A.~P. Dawid, ``Conditional independence in statistical theory,'' \emph{Journal
  of the Royal Statistical Society: Series B (Methodological)}, vol.~41, no.~1,
  pp. 1--15, 1979.

\bibitem{spohn1980stochastic}
W.~Spohn, ``Stochastic independence, causal independence, and shieldability,''
  \emph{Journal of Philosophical logic}, vol.~9, no.~1, pp. 73--99, 1980.

\bibitem{mouchart1984note}
M.~Mouchart and J.-M. Rolin, ``A note on conditional independence,''
  \emph{Statistica}, vol.~44, p. 557, 1984.

\bibitem{pearl1987graphoids}
J.~Pearl and A.~Paz, ``Graphoids: a graph-based logic for reasoning about
  relevance relations,'' \emph{Advances in Artificial Intelligence}, pp.
  357--363, 1987.

\bibitem{studeny1989multiinformation}
M.~Studen{\'y}, ``Multiinformation and the problem of characterization of
  conditional independence relations,'' \emph{Problems of Control and
  Information Theory}, no.~18, pp. 3--16, 1989.

\bibitem{studeny1990conditional}
------, ``Conditional independence relations have no finite complete
  characterization,'' \emph{Information Theory, Statistical Decision Functions
  and Random Processes}, pp. 377--396, 1992.

\bibitem{geiger1991axioms}
D.~Geiger, A.~Paz, and J.~Pearl, ``Axioms and algorithms for inferences
  involving probabilistic independence,'' \emph{Information and Computation},
  vol.~91, no.~1, pp. 128--141, 1991.

\bibitem{geiger1993logical}
D.~Geiger and J.~Pearl, ``Logical and algorithmic properties of conditional
  independence and graphical models,'' \emph{The Annals of Statistics}, pp.
  2001--2021, 1993.

\bibitem{geiger1999quantifier}
D.~Geiger and C.~Meek, ``Quantifier elimination for statistical problems,'' in
  \emph{Proceedings of the Fifteenth conference on Uncertainty in artificial
  intelligence}, 1999, pp. 226--235.

\bibitem{niepert2012logical}
M.~Niepert, ``Logical inference algorithms and matrix representations for
  probabilistic conditional independence,'' \emph{arXiv preprint
  arXiv:1205.2621}, 2012.

\bibitem{gyssens2014completeness}
M.~Gyssens, M.~Niepert, and D.~Van~Gucht, ``On the completeness of the
  semigraphoid axioms for deriving arbitrary from saturated conditional
  independence statements,'' \emph{Information Processing Letters}, vol. 114,
  no.~11, pp. 628--633, 2014.

\bibitem{hannula2019facets}
M.~Hannula, {\AA}.~Hirvonen, J.~Kontinen, V.~Kulikov, and J.~Virtema, ``Facets
  of distribution identities in probabilistic team semantics,'' in
  \emph{European Conference on Logics in Artificial Intelligence}.\hskip 1em
  plus 0.5em minus 0.4em\relax Springer, 2019, pp. 304--320.

\bibitem{khamis2020decision}
M.~A. Khamis, P.~G. Kolaitis, H.~Q. Ngo, and D.~Suciu, ``Decision problems in
  information theory,'' \emph{arXiv preprint arXiv:2004.08783}, 2020.

\bibitem{li2021undecidability}
C.~T. Li, ``The undecidability of conditional affine information inequalities
  and conditional independence implication with a binary constraint,'' in
  \emph{2021 IEEE Information Theory Workshop}, 2021.

\bibitem{li2021first}
------, ``First-order theory of probabilistic independence and single-letter
  characterizations of capacity regions,'' \emph{arXiv preprint
  arXiv:2108.07324}, 2021.

\bibitem{wong2000implication}
S.~M. Wong, C.~J. Butz, and D.~Wu, ``On the implication problem for
  probabilistic conditional independency,'' \emph{IEEE Transactions on Systems,
  Man, and Cybernetics-Part A: Systems and Humans}, vol.~30, no.~6, pp.
  785--805, 2000.

\bibitem{niepert2010logical}
M.~Niepert, D.~Van~Gucht, and M.~Gyssens, ``Logical and algorithmic properties
  of stable conditional independence,'' \emph{International Journal of
  Approximate Reasoning}, vol.~51, no.~5, pp. 531--543, 2010.

\bibitem{niepert2013conditional}
M.~Niepert, M.~Gyssens, B.~Sayrafi, and D.~Van~Gucht, ``On the conditional
  independence implication problem: A lattice-theoretic approach,''
  \emph{Artificial Intelligence}, vol. 202, pp. 29--51, 2013.

\bibitem{koehler2014saturated}
H.~Koehler and S.~Link, ``Saturated conditional independence with fixed and
  undetermined sets of incomplete random variables,'' in \emph{Proceedings of
  the Thirtieth Conference on Uncertainty in Artificial Intelligence}, 2014,
  pp. 410--419.

\bibitem{zhang1997non}
Z.~Zhang and R.~W. Yeung, ``A non-{S}hannon-type conditional inequality of
  information quantities,'' \emph{IEEE Trans. Inf. Theory}, vol.~43, no.~6, pp.
  1982--1986, 1997.

\bibitem{yeung1997framework}
R.~W. Yeung, ``A framework for linear information inequalities,'' \emph{IEEE
  Trans. Inf. Theory}, vol.~43, no.~6, pp. 1924--1934, 1997.

\bibitem{zhang1998characterization}
Z.~Zhang and R.~W. Yeung, ``On characterization of entropy function via
  information inequalities,'' \emph{IEEE Trans. Inf. Theory}, vol.~44, no.~4,
  pp. 1440--1452, 1998.

\bibitem{yeung2008information}
R.~W. Yeung, \emph{Information theory and network coding}.\hskip 1em plus 0.5em
  minus 0.4em\relax Springer Science \& Business Media, 2008.

\bibitem{chan2008dualities}
T.~Chan and A.~Grant, ``Dualities between entropy functions and network
  codes,'' \emph{IEEE Trans. Inf. Theory}, vol.~54, no.~10, pp. 4470--4487,
  2008.

\bibitem{yan2012implicit}
X.~Yan, R.~W. Yeung, and Z.~Zhang, ``An implicit characterization of the
  achievable rate region for acyclic multisource multisink network coding,''
  \emph{IEEE Trans. Inf. Theory}, vol.~58, no.~9, pp. 5625--5639, 2012.

\bibitem{makarychev2002new}
K.~Makarychev, Y.~Makarychev, A.~Romashchenko, and N.~Vereshchagin, ``A new
  class of non-{S}hannon-type inequalities for entropies,''
  \emph{Communications in Information and Systems}, vol.~2, no.~2, pp.
  147--166, 2002.

\bibitem{dougherty2006six}
R.~Dougherty, C.~Freiling, and K.~Zeger, ``Six new non-{S}hannon information
  inequalities,'' in \emph{2006 IEEE ISIT}.\hskip 1em plus 0.5em minus
  0.4em\relax IEEE, Jul 2006, pp. 233--236.

\bibitem{matus2007infinitely}
F.~Mat{\'u}{\v{s}}, ``Infinitely many information inequalities,'' in \emph{2007
  IEEE ISIT}.\hskip 1em plus 0.5em minus 0.4em\relax IEEE, Jun 2007, pp.
  41--44.

\bibitem{xu2008projection}
W.~Xu, J.~Wang, and J.~Sun, ``A projection method for derivation of
  non-{S}hannon-type information inequalities,'' in \emph{2008 IEEE
  ISIT}.\hskip 1em plus 0.5em minus 0.4em\relax IEEE, 2008, pp. 2116--2120.

\bibitem{dougherty2011non}
R.~Dougherty, C.~Freiling, and K.~Zeger, ``Non-{S}hannon information
  inequalities in four random variables,'' \emph{arXiv preprint
  arXiv:1104.3602}, 2011.

\bibitem{kaced2013conditional}
T.~Kaced and A.~Romashchenko, ``Conditional information inequalities for
  entropic and almost entropic points,'' \emph{IEEE Trans. Inf. Theory},
  vol.~59, no.~11, pp. 7149--7167, 2013.

\bibitem{yeung1996itip}
\BIBentryALTinterwordspacing
R.~W. Yeung and Y.~O. Yan, ``{ITIP} - information theoretic inequality
  prover,'' 1996. [Online]. Available:
  \url{http://user-www.ie.cuhk.edu.hk/~ITIP/}
\BIBentrySTDinterwordspacing

\bibitem{gurpinar2019use}
E.~G{\"u}rp{\i}nar and A.~Romashchenko, ``How to use undiscovered information
  inequalities: Direct applications of the copy lemma,'' in \emph{2019 IEEE
  ISIT}.\hskip 1em plus 0.5em minus 0.4em\relax IEEE, 2019, pp. 1377--1381.

\bibitem{ho2020proving}
S.~W. {Ho}, L.~{Ling}, C.~W. {Tan}, and R.~W. {Yeung}, ``Proving and disproving
  information inequalities: Theory and scalable algorithms,'' \emph{IEEE Trans.
  Inf. Theory}, vol.~66, no.~9, pp. 5522--5536, 2020.

\bibitem{li2021automatedisit}
C.~T. Li, ``An automated theorem proving framework for information-theoretic
  results,'' in \emph{2021 IEEE International Symposium on Information Theory
  (ISIT)}, 2021, pp. 2750--2755.

\bibitem{gomez2017defining}
A.~G{\'o}mez, C.~Mejia, and J.~A. Montoya, ``Defining the almost-entropic
  regions by algebraic inequalities,'' \emph{International Journal of
  Information and Coding Theory}, vol.~4, no.~1, pp. 1--18, 2017.

\bibitem{khamis2021bag}
M.~A. Khamis, P.~G. Kolaitis, H.~Q. Ngo, and D.~Suciu, ``Bag query containment
  and information theory,'' \emph{ACM Transactions on Database Systems (TODS)},
  vol.~46, no.~3, pp. 1--39, 2021.

\bibitem{yeung2021machine}
R.~W. Yeung and C.~T. Li, ``Machine-proof of entropy inequalities,'' \emph{IEEE
  BITS the Information Theory Magazine}, 2021.

\bibitem{herrmann1995undecidability}
C.~Herrmann, ``On the undecidability of implications between embedded
  multivalued database dependencies,'' \emph{Information and Computation}, vol.
  122, no.~2, pp. 221--235, 1995.

\bibitem{herrmann2006corrigendum}
------, ``Corrigendum to "on the undecidability of implications between
  embedded multivalued database dependencies"[inform. and comput. 122 (1995)
  221--235],'' \emph{Information and Computation}, vol. 204, no.~12, pp.
  1847--1851, 2006.

\bibitem{fagin1977multivalued}
R.~Fagin, ``Multivalued dependencies and a new normal form for relational
  databases,'' \emph{ACM Transactions on Database Systems (TODS)}, vol.~2,
  no.~3, pp. 262--278, 1977.

\bibitem{dougherty2006unachievability}
R.~Dougherty, C.~Freiling, and K.~Zeger, ``Unachievability of network coding
  capacity,'' \emph{IEEE Transactions on Information Theory}, vol.~52, no.~6,
  pp. 2365--2372, 2006.

\bibitem{lehman2005networkmodel}
A.~R. Lehman and E.~Lehman, ``Network coding: Does the model need tuning?'' in
  \emph{SODA}, vol.~5, 2005, pp. 499--504.

\bibitem{li2004network}
Z.~Li and B.~Li, ``Network coding: The case of multiple unicast sessions,'' in
  \emph{Allerton Conference on Communications}, vol.~16, no.~8, 2004.

\bibitem{dougherty2006nonreversibility}
R.~Dougherty and K.~Zeger, ``Nonreversibility and equivalent constructions of
  multiple-unicast networks,'' \emph{IEEE Transactions on Information Theory},
  vol.~52, no.~11, pp. 5067--5077, 2006.

\bibitem{bar2011index}
Z.~Bar-Yossef, Y.~Birk, T.~Jayram, and T.~Kol, ``Index coding with side
  information,'' \emph{IEEE Transactions on Information Theory}, vol.~57,
  no.~3, pp. 1479--1494, 2011.

\bibitem{lubetzky2009nonlinear}
E.~Lubetzky and U.~Stav, ``Nonlinear index coding outperforming the linear
  optimum,'' \emph{IEEE Transactions on Information Theory}, vol.~55, no.~8,
  pp. 3544--3551, 2009.

\bibitem{el2010index}
S.~El~Rouayheb, A.~Sprintson, and C.~Georghiades, ``On the index coding problem
  and its relation to network coding and matroid theory,'' \emph{IEEE
  Transactions on Information Theory}, vol.~56, no.~7, pp. 3187--3195, 2010.

\bibitem{effros2015equivalence}
M.~Effros, S.~El~Rouayheb, and M.~Langberg, ``An equivalence between network
  coding and index coding,'' \emph{IEEE Transactions on Information Theory},
  vol.~61, no.~5, pp. 2478--2487, 2015.

\bibitem{slobodskoi1981undecidability}
A.~M. Slobodskoi, ``Undecidability of the universal theory of finite groups,''
  \emph{Algebra i logika}, vol.~20, no.~2, pp. 207--230, 1981.

\bibitem{keedwell2015latin}
A.~D. Keedwell and J.~D{\'e}nes, \emph{Latin squares and their
  applications}.\hskip 1em plus 0.5em minus 0.4em\relax Elsevier, 2015.

\bibitem{oxley2006matroid}
J.~G. Oxley, \emph{Matroid theory}.\hskip 1em plus 0.5em minus 0.4em\relax
  Oxford University Press, USA, 2006, vol.~3.

\bibitem{dougherty2007networks}
R.~Dougherty, C.~Freiling, and K.~Zeger, ``Networks, matroids, and
  non-{S}hannon information inequalities,'' \emph{IEEE Transactions on
  Information Theory}, vol.~53, no.~6, pp. 1949--1969, 2007.

\bibitem{herrmann1987frames}
C.~Herrmann, ``Frames of permuting equivalences,'' \emph{Acta Sci. Math},
  vol.~51, no. 1-2, pp. 93--101, 1987.

\bibitem{markov1951impossibility}
A.~Markov, ``Impossibility of certain algorithms in the theory of associative
  systems,'' \emph{Journal of Symbolic Logic}, vol.~16, no.~3, 1951.

\bibitem{post1947recursive}
E.~L. Post, ``Recursive unsolvability of a problem of {T}hue,'' \emph{The
  Journal of Symbolic Logic}, vol.~12, no.~1, pp. 1--11, 1947.

\bibitem{turing1950word}
A.~M. Turing, ``The word problem in semi-groups with cancellation,''
  \emph{Annals of Mathematics}, pp. 491--505, 1950.

\bibitem{gurevich1966problem}
Y.~S. Gurevich, ``The problem of equality of words for certain classes of
  semigroups,'' \emph{Algebra i logika}, vol.~5, no.~5, pp. 25--35, 1966.

\bibitem{kurosh1963lectures}
A.~Kurosh, ``Lectures on general algebra,'' 1963.

\bibitem{li2021undecidabilityarxiv}
C.~T. Li, ``The undecidability of conditional affine information inequalities
  and conditional independence implication with a binary constraint,''
  \emph{arXiv preprint arXiv:2104.05634}, 2021.

\bibitem{fulton2013representation}
W.~Fulton and J.~Harris, \emph{Representation theory: a first course}.\hskip
  1em plus 0.5em minus 0.4em\relax Springer Science \& Business Media, 2013,
  vol. 129.

\bibitem{dougherty2005insufficiency}
R.~Dougherty, C.~Freiling, and K.~Zeger, ``Insufficiency of linear coding in
  network information flow,'' \emph{IEEE transactions on information theory},
  vol.~51, no.~8, pp. 2745--2759, 2005.

\end{thebibliography}

\end{document}